\documentclass[fleqn,usenatbib]{mnras}

\usepackage[T1]{fontenc}
\usepackage{ae,aecompl}

\usepackage{amssymb}	% Extra maths symbols
\usepackage{epsfig}
\usepackage{aas_macros}
\usepackage{natbib}
\usepackage{times}
\usepackage{graphicx, bm, amssymb, xcolor}
\usepackage{verbatim}
\usepackage[all]{hypcap}
\usepackage{xspace}
\usepackage{multirow}
\usepackage[fleqn]{amsmath}	% Advanced maths commands
\usepackage{multirow}
\usepackage{colortbl}

\newcommand{\msun}{\mbox{M$_\odot$}}
\newcommand{\yr}{\mbox{${\rm yr}$}}
\newcommand{\myr}{\mbox{${\rm Myr}$}}
\newcommand{\gyr}{\mbox{${\rm Gyr}$}}
\newcommand{\pc}{\mbox{${\rm pc}$}}
\newcommand{\kpc}{\mbox{${\rm kpc}$}}
\newcommand{\kms}{\mbox{${\rm km}~{\rm s}^{-1}$}}

\title[Age distributions of cluster populations]{A tight relation between the age distributions of stellar clusters and the properties of the interstellar medium in the host galaxy}

\author[M.~Miholics, J.~M.~D.~Kruijssen \& A.~Sills]{Meghan Miholics$^{1,2}$\thanks{miholim@mcmaster.ca}, J.~M.~Diederik Kruijssen$^2$\thanks{kruijssen@uni-heidelberg.de} and Alison Sills$^1$\thanks{asills@mcmaster.ca}
\\
$^{1}$Department of Physics \& Astronomy, McMaster University, 1280 Main Street West, Hamilton ON, L8S 4M1 Canada\\
$^{2}$Astronomisches Rechen-Institut, Zentrum f\"{u}r Astronomie der Universit\"{a}t Heidelberg, M\"{o}nchhofstra\ss e 12-14, 69120 Heidelberg, Germany}

\date{Accepted 2017 May 24. Received 2017 May 24; in original form 2017 February 22.}

\pubyear{2017}

\begin{document}
\label{firstpage}
\pagerange{\pageref{firstpage}--\pageref{lastpage}}
\maketitle

\begin{abstract}
The age distributions of stellar cluster populations have long been proposed to probe the recent formation history of the host galaxy. However, progress is hampered by the limited understanding of cluster disruption by evaporation and tidal shocks. We study the age distributions of clusters in smoothed particle hydrodynamics simulations of isolated disc galaxies, which include a self-consistent, physical model for the formation and dynamical evolution of the cluster population and account for the variation of cluster disruption in time and space. We show that the downward slope of the cluster age distribution due to disruption cannot be reproduced with a single functional form, because the disruption rate exhibits systematic trends with cluster age (the `cruel cradle effect'). This problem is resolved by using the median cluster age to trace cluster disruption. Across 120 independent galaxy snapshots and simulated cluster populations, we perform two-dimensional power law fits of the median cluster age to various macroscopic physical quantities and find that it scales as $t_{\rm med}\propto \Sigma^{-0.51\pm0.03}\sigma_{\rm 1D}^{-0.85\pm0.10}M_{\rm min}^\gamma$, for the gas surface density $\Sigma$, gas velocity dispersion $\sigma_{\rm 1D}$, and minimum cluster mass $M_{\rm min}$. This scaling accurately describes observed cluster populations and indicates disruption by impulsive tidal shocks from the interstellar medium. The term $M_{\rm min}^\gamma$ provides a model-independent way to measure the mass dependence of the cluster disruption time $\gamma$. Finally, the ensemble-average cluster lifetime depends on the gas density less strongly than the instantaneous disruption time of single clusters. These results reflect the variation of cluster disruption in time and space. We provide quantitative ways of accounting for these physics in cluster population studies.
\end{abstract}


%%%%%%%%%%%%%%%%% BODY OF PAPER %%%%%%%%%%%%%%%%%%

\section{Introduction}
Understanding the formation and evolution of galaxies over cosmic history is one of the main goals of modern astrophysics \citep[e.g.][]{White91,Cole94,Sanders96,Kennicutt98,Cole00,Conselice03,vanDokkum05,McConnachie09,Hopkins10,Vogelsberger14,Schaye15}. Since all but the smallest galaxies in the Universe host populations of star clusters \citep{Harris13} and the lifetimes of the most massive star clusters (globular clusters) are in excess of $10~\gyr$ \citep{Marin09}, star clusters have incredible potential to trace the formation and evolution of galaxies \citep{Searle78,Ashman92,Schweizer98,Larsen01,Bastian05,Glatt10,Kruijssen15b,Caldwell16}. However, the promise of tracing galaxy formation and evolution with stellar clusters can only be fulfilled if a systematic census is obtained of the demographics of extragalactic cluster populations, and if a theoretical framework is constructed to connect these demographics to the formation histories of galaxies. While a wide range of studies and ongoing surveys have driven major progress in addressing the first of these requirements \citep[e.g.][]{Jordan07,Bastian12,Dalcanton12,Adamo15,Calzetti15}, the connection to galaxy evolution is still in its early development \citep[e.g.][]{Kravtsov05,Kruijssen11,Tonini13,Li14,Kruijssen15b}.

The potential to use star clusters as tracers for galactic evolution becomes especially clear when considering the appreciable impact that the host galaxy has on the formation and evolution of individual star clusters. Star clusters are formed from the cold dense molecular hydrogen in the galaxy \citep[e.g.][]{Longmore14}. However, they are also destroyed over time through gravitational interactions with the host galaxy such as tidally-limited evaporation and impulsive shocks \citep[e.g.][]{Spitzer87}. The rate of cluster destruction also increases due to the substructure in the interstellar medium (ISM), which produces rapid and strong variations in the tidal field strength and leads to more tidal heating than in isolation \citep{Lamers06,Kruijssen11}. Subsequently, the properties of the star cluster population as a whole, such as the age and mass distributions, will be intrinsically linked to the host galaxy evolution. However, to use the age and mass distributions of star clusters as probes of galactic evolution, we must first have a systematic understanding of how galactic evolution effects these properties. The effects of the tidal field on individual star clusters has been studied extensively in the literature \citep{Vesperini97,Baumgardt03,Gieles06,Rieder13,Webb14a,Webb14b,Renaud15b}. However, these studies have rarely followed the self-consistent, simultaneous evolution of stellar clusters and their host galaxy. These works also typically focus on the evolution of only a few clusters at a time rather than of an entire population. In order to trace galaxy evolution using stellar clusters, we need to make use of the full population.

Previous studies have attempted to model the age and mass distributions of star cluster populations and compare them to observations in several galaxies \citep[e.g.][]{Boutloukos03,Hunter03,Gieles05,Lamers05,Smith07,Larsen09,Bastian12}. These studies have shown that the disruption rate of star clusters varies from galaxy to galaxy and in particular that galaxies with the highest star formation rates also have the highest disruption rates. However, these studies make a variety of the simplifying assumptions, including that the star and cluster formation rate and the cluster disruption rate are constant in time and throughout the galaxy, even though the cluster age distribution scales linearly with the cluster formation rate at each age. Although these studies represent an excellent first step in this field, the theories of star formation and star cluster disruption demonstrate that both are dependent on a number of environmental factors such as ambient density and pressure. Hence, the assumption that star formation and cluster disruption is constant through a galaxy is not valid even for an individual star cluster since its orbit around the galaxy may take it through areas of varying density. The assumption that these rates are constant in time is especially inaccurate when considering transient events such as galaxy mergers or tidal interactions.

\subsection{Preliminary work} \label{sec:prelim}
We must understand the impact of variable star formation and cluster disruption rates on the observed properties of star cluster populations to fully unlock the potential for using star clusters as tracers of galactic evolution. A first step toward understanding the interplay between cluster population properties and the evolution of galaxies was undertaken by \cite{Kruijssen11,Kruijssen12}. This work was based on Smoothed Particle Hydrodynamics (SPH) simulations of both isolated disc galaxies and galaxy mergers, coupled to semi-analytical models for star cluster formation and destruction. Star clusters are formed subgrid within new star particles. The dynamical mass loss of all clusters is then followed throughout the simulation by measuring the tidal field tensors at each timestep and predicting the corresponding mass loss rates due to tidal evaporation and tidal heating.

These galaxy simulations provided a number of new insights regarding the interaction between galaxies and their star cluster populations. In their isolated disc galaxy simulations, \citet{Kruijssen11} found that up to 90 per cent of cluster disruption is caused by tidal heating due to encounters with giant molecular clouds (GMCs). Since the rate of encounters is increases with the gas density, the distribution of gas plays a large role in shaping the observable properties of the star cluster population. For instance, the mean age of clusters in isolated disc galaxies was found to be anti-correlated with the gas density, indicating higher disruption rates in high-density environments. Most importantly, \citet{Kruijssen11} identified two mechanisms that together lead to a disruption rate that decreases with increasing cluster age (later dubbed the `cruel cradle effect' in \citealt{Kruijssen12b}, also see \citealt{Elmegreen10}). Both mechanisms are related to the fact that clusters are born in environments of high gas density. The first mechanism, `natural selection', states that clusters formed in less destructive environments survive over longer time-scales, implying that the disruption rate of the surviving population decreases as the population ages. The second mechanism, `cluster migration', states that cluster form near the density peaks in the ISM, which lead to an initial phase of rapid tidal destruction that fades as galactic dynamics allow the clusters to migrate away from their birth sites. These two effects leave a distinct imprint on the observed cluster age distribution at old ages. As was shown by \citet{Lamers05}, the cluster age distribution has a negative slope due to ongoing cluster disruption. However, the fact that the disruption rate decreases with increasing age acts to make the slope of the distribution shallower, in a way that was predicted by \citet{Kruijssen11} to trace the density contrast (and by proxy the absolute ISM density) in the host galaxy during the first $\sim100~\myr$ after cluster formation.

\citet{Kruijssen12} focussed on analysing the galaxy merger models in the suite of simulations and showed that both cluster formation and destruction are enhanced in galaxy mergers, with the latter process eventually dominating and leading to a decrease of the number of clusters. It was also found that galaxy mergers significantly affect the mass distribution of clusters, which forms a peak due to the preferential destruction of low mass clusters. Additionally, galaxy mergers were shown to leave a distinct imprint on the age distribution of star clusters by \citet{Kruijssen11}, where it was found that although there is a significant increase in star formation just after the final coalescence of the merging galaxies begins, no clusters born during the time of coalescence survive due to the high densities and pressures present during the merger. This results in a gap in the cluster age distribution (in particular at low masses) that reflects when the galaxy underwent a merger. These results illustrate how current models are starting to enable the tracing of galaxy evolution by using cluster populations.

\subsection{This paper}
The initial results discussed above have demonstrated that cluster formation and disruption are environmentally-dependent processes, which is in qualitative agreement with recent observational results \citep{Lamers05b,Bastian12,Fouesneau14,Adamo15b}. Leading on from these results, the goal of the present work is to extend the analysis of the simulations to carry out a systematic study of how the star cluster age distributions depend on local galactic conditions such as the gas density and gas velocity dispersion. In this way, we can provide expected relationships between observed star cluster properties and galactic properties that can be easily tested by observations. With the arrival of the Atacama Large Millimeter/submillimeter Array (ALMA), it is now possible to test current ideas about the interaction between the cluster population and the ISM at unprecedented sensitivity and resolution \citep[e.g.][]{Freeman17}.

To achieve these goals we use the isolated disc galaxy models from \citet{Kruijssen11,Kruijssen12}, which have properties that are approximately constant in time and thus provide a controlled environment to probe the connection between the host galaxy properties and the resulting cluster age distribution. In Section~\ref{sec:sims}, we first summarise the setup of the numerical simulations and the physics included in these models. We present the resulting cluster age distributions in Section~\ref{sec:agedis} and show that the median age is a good proxy for the disruption rate. In Section~\ref{sec:median}, we correlate the median cluster age to the host galaxy properties and demonstrate a close relation to the gas surface density and the gas velocity dispersion. This result is discussed in Section~\ref{sec:disc} and compared to observations of cluster populations in nearby galaxies. The conclusions of this paper are presented in Section~\ref{sec:concl}.

\section{Simulations} \label{sec:sims}
In this section, we will summarise the main components of the simulations and models used in this work. For a complete description, please refer to \citet{Kruijssen11,Kruijssen12}.

\subsection{Galaxy models}
\begin{table*}
\centering
\begin{minipage}{150mm}
\caption{Initial conditions for the considered disc galaxy models}\label{tab:discs}
\begin{tabular}{c c c c c c c c c c c c}
\hline
ID & $f_{\rm gas}$ & ${M_{\rm vir}}^a$ & $z$ & $\lambda$ & $N_{\rm halo}$ & $N_{\rm gas}$ & $N_{\rm disc}^{\rm star}$ & $N_{\rm bulge}^{\rm star}$ & ${M_{\rm part}^{\rm halo}}^a$ & ${M_{\rm part}^{\rm bary}}^a$ & Comments
\\\hline
1dA    &      0.20     &     10$^{12}$     &   2   &     0.05     &     10$^6$     &     10250     &     41000    &      10000   & $10^6$ & $8\times10^5$ &     low gas fraction      \\
1dB    &      0.30     &     10$^{12}$     &   2   &     0.05     &     10$^6$     &     15375     &     35875    &      10000   & $10^6$ & $8\times10^5$ &  standard model \\
1dC    &      0.50     &     10$^{12}$     &   2   &     0.05     &     10$^6$     &     25625     &     25625    &      10000   & $10^6$ & $8\times10^5$ &     high gas fraction      \\
1dD    &      0.30     &     5$\times10^{11}$ &   2   &       0.05   &       $5\times10^5$   &       7688   &       17938  &        5000 & $10^6$ & $8\times10^5$ &     half mass \\
1dE    &      0.30     &     10$^{12}$     &   2   &     0.05     &     10$^6$     &     15375     &     35875      &  0         & $10^6$ & $8\times10^5$ &     no bulge           \\
1dF    &      0.30     &     10$^{11}$     &   2   &     0.05     &     10$^6$     &     15375     &     35875    &      10000   & $10^5$ & $8\times10^4$ &     low mass           \\
1dG    &      0.30     &     10$^{12}$     &   2   &     0.10     &     10$^6$     &     15375     &     35875    &      10000   & $10^6$ & $8\times10^5$ &     high spin           \\
1dH    &      0.30     &     10$^{12}$     &   0   &     0.05     &     10$^6$     &     15375     &     35875    &      10000   & $10^6$ & $8\times10^5$ &     low concentration        \\
1dI    &      0.30     &     10$^{12}$     &   5   &     0.05     &     10$^6$     &     15375     &     35875    &      10000   & $10^6$ & $8\times10^5$ &     high concentration        \\
\hline
\end{tabular} \\
$^a$In solar masses ($\msun$).
\end{minipage}
\end{table*}
Each galaxy in the suite is comprised of the usual main components of a disc galaxy: a dark matter halo, a stellar and gaseous disc and a stellar bulge. The stellar and dark matter particles are evolved with the hydrodynamics code {\sc fi} \citep{Pelupessy05}, according to collisionless Newtonian dynamics using a Barnes-Hut tree method \citep{Barnes86}. The gravitational forces are softened on scales of approximately 200 pc. The gas particles are evolved according to SPH \citep{Monaghan92,Springel02}. A full description of the ISM model can be found in \cite{Pelupessy04,Pelupessy05}, but the main ingredients are as follows. The model solves for the ionisation and thermal balance in the neutral and ionised components of the ISM by including a number of relevant physical processes such as photoelectric heating from UV radiation in the optically-thin approximation, line cooling from 8 elements (H, He, C, N, O, Ne, Si and Fe), and cosmic ray ionisation. Star formation is included using a model for the gravitational instability of molecular clouds (see Section~\ref{sec:methods} below) and mechanical feedback from stellar winds and supernovae (SNe) is included using massless `pressure particles', which are assumed to contain 10 per cent of the SN output energy (i.e.~the remaining 90 per cent is radiated away, see e.g.~\citealt{Thornton98}). The use of pressure particles avoids the well-known overcooling problem and provides a more realistic injection of energy and momentum into the ISM than alternative solutions such as temporarily switching off cooling or dynamically decoupling feedback-affected particles.

These numerical ingredients result in stable galaxy models with gradually declining star formation rates \citep[SFRs, see e.g.][Figure~4]{Kruijssen11}, as expected for a closed-box setup of an isolated disc galaxy. Star formation in the galaxies reproduces the observed relation between the SFR (surface density) and the (molecular) gas mass (surface density) \citep{Schmidt59,Kennicutt98,Leroy08}, as discussed in \citet{Pelupessy09} and \citet{Krumholz11}, as well as the starburst duty cycle of dwarf galaxies \citep{Pelupessy04}. In addition, the interplay between star formation, feedback, heating, and cooling results in an ISM that is similarly structured to observations of the atomic and molecular phases in nearby galaxies, e.g.~in terms of the spatial power spectrum, the (metallicity-dependent) molecular gas fraction, and the disc scale height \citep{Pelupessy06}. These are all important ingredients when aiming to model ISM-driven cluster disruption.

When setting up the initial conditions of the galaxy models, four key parameters are chosen for each galaxy and can vary from galaxy to galaxy in the suite. These parameters are virial mass ($M_{\rm vir}$), gas fraction of the baryonic disc ($f_{\rm g}$), halo condensation redshift ($z$) and spin parameter ($\lambda =J|E|/GM_{\rm vir}^{5/2}$, where $J$ and $E$ are the angular momentum and energy of the halo, respectively). Table~\ref{tab:discs} shows these parameters for the galaxies in the suite as well as the number of particles and particle masses used to represent each component of the galaxy. The other properties of each galaxy follow from these parameters. Each galaxy is initialised to be in self-gravitating equilibrium \citep{Springel05}. The density profile of each galaxy's dark matter halo is given by \cite{Hernquist90}. The concentration of each dark matter halo is calculated using the chosen redshift and mass of the galaxy according to \cite{Bullock01}. The virial velocity, $V_{\rm vir}$, is also calculated using total mass and redshift according to $M_{\rm vir} =  V_{\rm vir}^3/\left(10GH(z)\right)$ where $H(z)$ is the Hubble constant at redshift $z$. The density profile for the stellar disc is exponential. The radial scale length of the disc is calculated according to the amount of rotation set by the spin parameter $\lambda$ \citep{Mo98}. The vertical scale height of the disc and the scale length of the bulge are set to $1/5$ times the radial scale length of the disc. The masses of the disc and bulge are assigned such that the fractions of the total galaxy mass are $m_d = 0.041$ and $m_b = 0.008$, respectively.

We emphasise that these galaxy simulations are dated in many respects when compared to the recent, major advances in numerical techniques \citep[e.g.][]{Dehnen12,Hopkins13c,Hu14}, subgrid physics \citep[e.g.][]{Hopkins14,Pakmor16} and dynamic range \citep[e.g.][]{Vogelsberger14,Schaye15}. However, the \citet{Kruijssen11,Kruijssen12} simulations remain the only models to date that include a complete and self-consistent census of the entire cluster population in galaxies, in that they provide an honest description of the interaction between the ISM and cluster formation and disruption while covering the complete cluster mass spectrum. The present study is intended as a precursor to future work that extends these models and will include the recent numerical developments. Examples are the E-MOSAICS project (Pfeffer et al.~in prep.; Kruijssen et al.~in prep.), which applies the star cluster population modelling of the present paper to cosmological zoom-in simulations in the context of the EAGLE simulations \citep{Schaye15}, as well as the currently ongoing implementation of this method (Reina-Campos et al.~in prep.) in the moving-mesh code Arepo \citep{Springel10}. With the analysis carried out in this paper, we will be able to motivate these future simulations using state-of-the-art hydrodynamics and subgrid physics.

\subsection{Star cluster formation and destruction models}
\label{sec:methods}
We use the subgrid star cluster formation and evolution model for galaxy simulations MOSAICS\footnote{This acronym refers to MOdelling Star cluster population Assembly In Cosmological Simulations.} that was developed in \citet{Kruijssen11} and has since been expanded by \citet{Kruijssen12c}, \citet{Reinacampos17}, and Pfeffer et al.~(in prep.). Here we use the initial version of this model and briefly summarise the treatment of cluster formation and evolution.

\subsubsection{Star cluster formation}
Star formation is implemented via a simple Jeans mass ($M_{\rm J}$) criterion, such that a gas particle is converted into a star particle when the following inequality is satisfied:
\begin{equation}
M_{\rm J} = \frac{\pi\rho}{6}\left(\frac{\pi s^2}{G \rho}\right)^{3/2} < M_{\rm ref} ,
\end{equation}
where $G$ is the gravitational constant, $\rho$ is the local density, $s$ is the local sound speed, and $M_{\rm ref}=5\times10^6~\msun$ is a reference mass-scale chosen to be close to the high-mass end of the observed GMC mass spectrum \citep[e.g.][]{Bolatto08,Freeman17}. This criterion enables star formation in cold and dense gas. When a region of gas satisfies this criterion, stars are formed on a time-scale $\tau_{\rm sf}$ proportional to the local free-fall time:
 \begin{equation}
 \tau_{\rm sf} = f_{\rm sf}t_{\rm ff} = \frac{f_{\rm sf}}{\sqrt{4\pi G\rho}} ,
 \end{equation}
where $f_{\rm sf}\approx10$. The process of forming stars is done by converting the gas particles identified as unstable to star particles stochastically with a probability $1 - \exp{(-{\rm d}t/\tau_{\rm sf})}$. The effects of stellar winds and supernovae are included via mechanical heating of the ISM \citep{Pelupessy04,Pelupessy05}.

Each time a gas particle is converted to a star particle, the formation of star clusters is implemented subgrid by generating a number of star clusters within the star particle. The cluster formation efficiency (CFE), i.e.~the fraction of star formation occurring in bound clusters is set to 90 per cent, implying that 10 per cent of the star particle mass is immediately assigned to the field. This value is much higher than predicted by theory or observed in real galaxies \citep{Kruijssen12c,Adamo15,Johnson16}, but for the purpose of this study (which does not deal with the absolute number of clusters) the precise value of the CFE is not important -- the use of a single value effectively corresponds to the assumption that the CFE is constant. The clusters are assigned masses such that they follow a power-law distribution with an exponential truncation \citep{Schechter76}:
\begin{equation}
N {\rm d}M \propto M^{-\alpha}\exp{(-M/M_{*})}{\rm d}M ,
\end{equation}
where $N$ is the number of clusters, $M$ is the cluster mass, $\alpha=2$ is the power law slope, and $M_{*}=2.5 \times 10^6~\msun$ is the truncation mass. The maximum cluster mass formed in original version of MOSAICS cannot exceed the particle mass (about $10^{5.8}~\msun$ for most simulations in the suite). This restriction is omitted in future applications of the model, for which a new stochastic sampling algorithm has been developed (Pfeffer et al.~in prep.). The minimum initial cluster mass is set to $10^2 M_{\odot}$. All clusters are given a stellar population according to a \citet{Kroupa01} initial mass fraction (IMF) with a minimum mass of 0.08 $M_{\odot}$ and a maximum mass corresponding to the most massive star at an age of $10^{6.6}$ yr (the minimum age in the Padova stellar evolution models used by the code, see \citealt{Marigo08}). We note that the star clusters and field stars do not physically disperse in the simulation but remain within the star particle. For each particle, the code keeps track of how much mass is in field stars versus clusters. The ratio of field stars to clusters changes throughout the simulation since star clusters lose mass, which in turn creates more field stars.

\subsubsection{Star cluster destruction}
Star cluster destruction is implemented in MOSAICS \citep{Kruijssen11} by calculating two mass loss rates for each cluster, i.e.~the mass loss by stellar evolution from the Padova models \citep{Marigo08} and dynamical evolution. The dynamical mass loss of the clusters also accounts for stellar-mass dependent escape rates \citep{Kruijssen08,Kruijssen09}, but the output from this module is not used in the present work, because we only focus on cluster masses and ages. Two processes responsible for mass loss via dynamical evolution are accounted for. The first mechanism is gravitational interactions between stars, known as two-body relaxation, which causes stars to move outwards and evaporate over the tidal boundary, removing them from the cluster \citep{Spitzer87,Heggie03}. The second mechanism is tidal shocking whereby stars gain energy through rapid variations in the tidal field caused by substructure such as GMCs or spiral arms \citep{Spitzer87,Kundic95,Gieles06}.

The mass loss rate for a cluster of mass $M$ is calculated as:
\begin{equation}
\left(\frac{{\rm d}M}{{\rm d}t}\right)_{\rm dis} = \left(\frac{{\rm d}M}{{\rm d}t}\right)_{\rm rlx} + \left(\frac{{\rm d}M}{{\rm d}t}\right)_{\rm sh} = -\frac{M}{t_{\rm rlx}} -\frac{M}{t_{\rm sh}} ,
\end{equation}
where $t_{\rm rlx}$ and $t_{\rm sh}$ are the time-scales for the complete dissolution of the cluster via two-body relaxation and tidal shocks respectively. The calculation of these time-scales requires knowledge of the local tidal field acting on each cluster. To this end, the tidal tensor at the position of the clusters is calculated. The tidal tensor is the matrix of second derivatives of the galactic potential, $\Phi$:
\begin{equation}
T_{ij} = - \frac{\partial^2 \Phi}{\partial x_i \partial x_j} .
\end{equation}
The tidal tensor is obtained by numerical differentiation of the force field at the position of each star particle that contains clusters.

The full derivation of $t_{\rm rlx}$ and $t_{\rm sh}$ can be found in \citet{Kruijssen11}. The final expression obtained for $t_{\rm rlx}$ is the following:
\begin{equation}
t_{\rm rlx} = 1.7~\gyr \left(\frac{M}{10^4~\msun}\right)^{\gamma} \left(\frac{T}{10^4~\gyr^{-2}}\right)^{-1/2} ,
\end{equation}
where $\gamma$ represents the mass-dependence of the disruption time-scales (see Section~\ref{sec:derivation} for more details) and $T$ is the tidal field strength experienced by the cluster. The tidal field strength is found by taking the maximum eigenvalue of the tidal tensor. In the simulations, $\gamma = 0.62$, which is broadly consistent with $N$-body simulations performed by \citep{Baumgardt03} and observations of clusters in multiple galaxies \citep{Lamers05b}.

The final expression obtained for the disruption time-scale due to tidal shocks, $t_{\rm sh}$ is the following:
\begin{equation}
t_{\rm sh} = 3.1~\gyr \left(\frac{M}{10^4~\msun}\right) \left(\frac{r_{\rm h}}{\pc}\right)^{-3} \left(\frac{I_{\rm tid}}{10^4~\gyr^{-2}}\right)^{-1}\left(\frac{\Delta t}{\myr}\right) ,
\end{equation}
where $r_{\rm h}$ is the radius containing half the mass (half-mass radius), $I_{\rm tid}$ is the tidal heating parameter and $\Delta t$ is the time since the last tidal shock. The tidal heating parameter is given by the following \citep{Gnedin03,Prieto08}:
\begin{equation}
I_{\rm tid} = \sum_{i.j} \left(\int T_{ij}{\rm d}t\right)^2 A_{{\rm w},ij} ,
\end{equation}
where $A_{{\rm w},ij}$ is the adiabatic correction \citep{Weinberg94a,Weinberg94b,Weinberg94c}. This term accounts for the fact that some of the energy injected by tidal shocks is absorbed by the cluster through expansion rather than given to the stars as kinetic energy. The adiabatic correction gives the fraction of the energy injected by the tidal shock that is actually absorbed by the stars as kinetic energy. It is given by
\begin{equation}
A_{{\rm w},ij} = \left(1 + \overline{\omega}^2\tau_{ij}^2\right)^{-3/2} ,
\end{equation}
where $\overline{\omega}\propto M^{1/2}r_{\rm h}^{-3/2}$ is the average angular frequency of stars in the cluster (proportional to square root of the cluster density) and $\tau_{ij}$ is the shock time-scale in the corresponding component \citep{Gnedin99b}. Shocks are identified independently in each component of the tidal tensor, and are separated by minima that are no higher than 39 per cent of the preceding maxima. The adiabatic correction introduces an additional dependence of the shock time-scale on cluster mass and density. For a given radius, high mass clusters will have a higher angular frequency and hence a smaller adiabatic correction. Consequently, a high mass cluster absorbs more energy via expansion than a low mass cluster for the same shock and is more stable against tidal shocks. 

Since the time-scale for disruption via tidal shocks carries a dependence on the half mass radius, which is expected to change as the mass changes \citep{Gieles11}, it is important to update the radii of the clusters as they lose mass in the simulations. The adopted mass-radius relation is $r_{\rm h} = 4.35 \text{ pc } (M/10^4 M_{\odot})^{0.225}$ and is consistent with $N$-body simulations \citep{Baumgardt03} of tidally filling star clusters. Recently, \citet{Kruijssen15b} and \citet{Gieles16} independently argued for an almost identical revision of the mass-radius relation to an environmentally-dependent and somewhat shallower form. Currently ongoing tests of this change have shown that in the context of MOSAICS, only the most short-lived, low-mass clusters are affected (Pfeffer et al.~in prep.). The results presented here are therefore qualitatively consistent with these recent revisions to the mass-radius relation. A detailed quantitative discussion of the mass-radius relation and its effect on the simulated cluster populations considered here can be found in \cite{Kruijssen11,Kruijssen12}.

\section{Cluster age distributions} \label{sec:agedis}
In this section, we discuss the cluster age distributions as a function of galactocentric radius in our fiducial galaxy model (1dB), with the goal of setting the reference point for comparisons between the models in Table~\ref{tab:discs} and for probing the environmental dependence of the cluster age distribution. Previously, the environmental dependence of cluster disruption was modelled with a single disruption time-scale `parameter' ($t_0$) per galaxy, which was taken to be representative for that galaxy across the cluster lifetime \citep{Lamers05}. Because the disruption rate in the simulations exhibits systematic trends with cluster age (see Section~\ref{sec:prelim}), this description is no longer capable of reproducing the functional form of the age distribution. We therefore prefer using a non-parametric diagnostic to trace the rapidity of cluster disruption. We argue that the median cluster age is a suitable quantity and conclude this section with a derivation of the relation between the median cluster age and the disruption parameter that was used in previous work.

\subsection{The radial variation of the cluster age distribution in the fiducial galaxy model 1dB}
\label{sec:1dB}
The age distribution of star clusters in a galaxy is shaped by the two competing processes of cluster formation and disruption. The connection between the age distribution and cluster disruption was demonstrated by \citet{Lamers05} who considered a simple model for the disruption of star clusters, taking the cluster disruption time as follows:
\begin{equation}
t_{\rm dis} = t_0\left(\frac{M}{M_{\odot}}\right)^{\gamma} ,
\end{equation}
where the disruption parameter $t_0$ is a constant depending on the tidal field experienced by the star cluster and $M$ is the cluster mass. Age distributions (normalised by cluster formation rate) calculated for these models are flat at young ages and slope downward at old ages \citep[see][Figure 5]{Lamers05}. The flattening at young ages occurs because a constant number of clusters are produced per unit time and no clusters are old enough to have been destroyed. The location of the `knee' in the distribution, where the slope becomes negative, corresponds to the disruption time of the shortest-lived (i.e.~lowest mass) cluster. Beyond the knee, higher mass clusters become disrupted and so fewer and fewer are left. In the \citet{Lamers05} model, the slope of the distribution at the old end is given by $-1/\gamma$ (assuming a power law initial cluster mass function with slope $-2$) and the age distribution thus also indicates the mass dependence of the disruption time-scale. A shallow slope of the age distribution indicates that the disruption time is strongly dependent on mass (i.e.~$\gamma$ is large). Therefore, age distributions are key tools for understanding how star cluster disruption depends on the local environment.

\begin{figure}
	\includegraphics[width=\hsize]{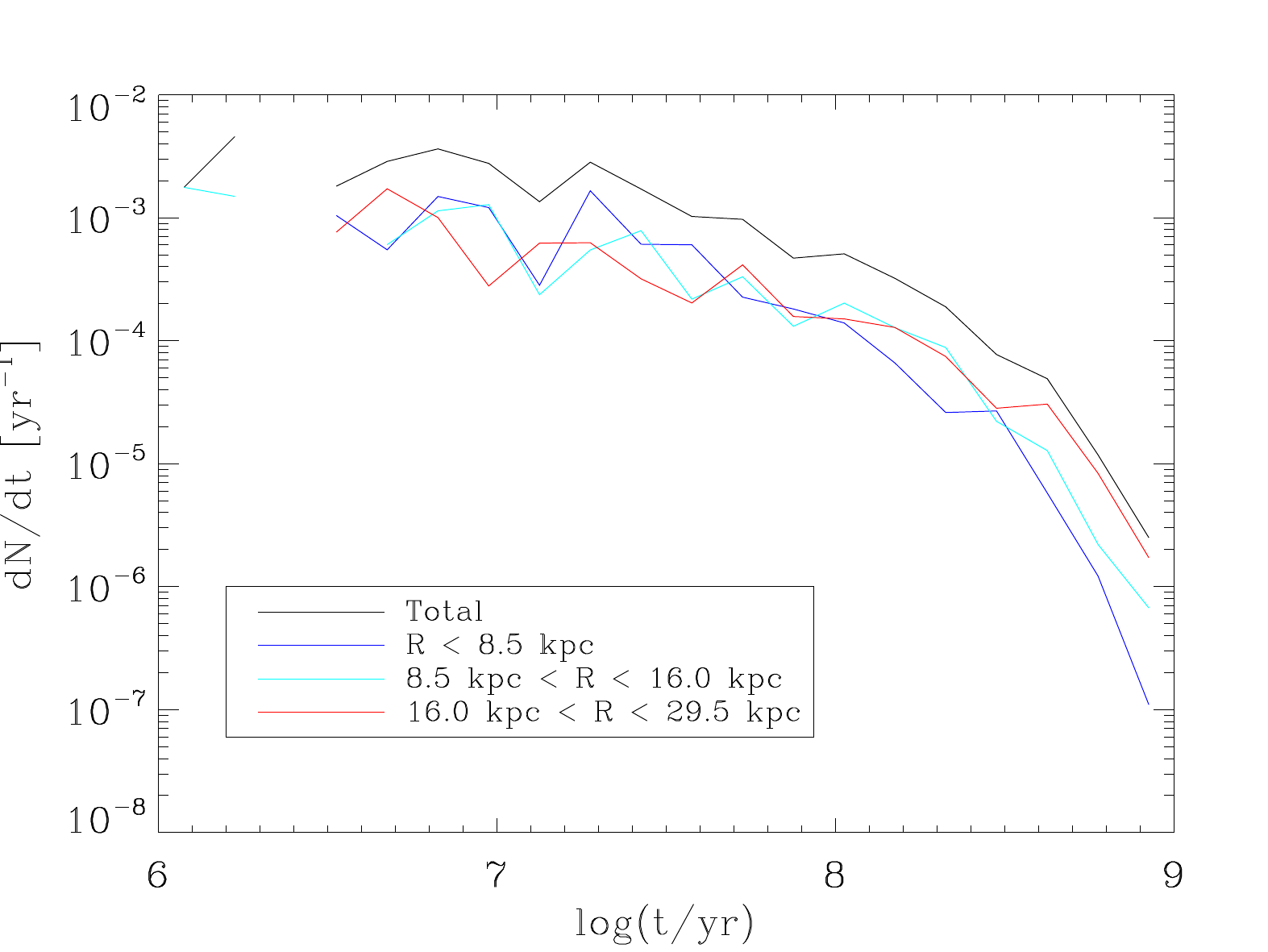}
    \caption{Age distributions of star clusters in a single snapshot of galaxy model 1dB. The total age distribution is shown in black, whereas the age distributions for clusters in the inner, middle and outer regions of the galaxy are shown in blue, cyan, and red respectively. Clusters are binned radially such that an equal number of clusters reside in each bin.}
    \label{fig:agedist}
\end{figure}
As a first step to understanding the environmental dependence of cluster disruption, we calculate the age distributions of star clusters in one of the galaxy simulations in the suite. We calculate an age distribution for clusters in three separate radial bins with boundaries defined such that an equal number of clusters occupies each of the bins. In Figure~\ref{fig:agedist}, we show an example of the these distributions for a single snapshot at $t=1.2~\gyr$ of our fiducial simulation 1dB. We also show the age distribution for all the clusters in the galaxy. The distributions all have roughly the same shape with a shallow slope at younger ages and a steep downturn at older ages. The absence of clusters with ages $<3~\myr$ arises due to small-number statistics, caused by using relatively high particle masses of $8\times10^5~\msun$. \citet{Adamo15b} compared the age distributions from \citet{Kruijssen11} to that observed for the cluster population in M83 and found that the overall shape provided an excellent fit to the data. In this particular example, we see that across the entire galaxy, 90 per cent of the cluster population has been lost after $\sim2\times10^8~\yr$ and 99 per cent has been lost after $\sim5\times10^8~\yr$. Howewer, clear differences exist amongst the radial bins. For example, the inner bin has a much steeper slope at ages in between $10^8$--$10^9$~yr than the outer bin, indicating that cluster disruption is more rapid in the inner regions of the galaxy than the outer regions. This behaviour is expected because the inner regions of the galaxy are richer in dense gas and thus we expect tidal shocks are more frequent and stronger.

Although it is clear that star cluster age distributions (and thus disruption) vary across the spatial extent of the galaxy, we would like to investigate which observable galactic properties correlate most strongly with the process of disruption. Before we can do so we first need a way to characterise the age distributions for comparison with one another and different galactic properties. In the \citet{Lamers05} model, the age distribution is completely characterised by $t_0$ (a timescale or parameter that characterises the strength of the tidal field) and $\gamma$ (the mass dependence of the disruption timescale). Since mass loss rates for each cluster are determined individually based on the tidal shocks it experiences, the disruption rate of clusters in a galaxy will vary with time and across the spatial extent of the galaxy (i.e. there is not a single value of $t_0$ valid everywhere in the galaxy).

It could be possible to account for the variation of the disruption time-scale if each of the constituent regions of the galaxy, as well as the galaxy as a whole, have a mean or effective $t_0$ for the population of clusters that we can then correlate with galactic properties. Fitting an age distribution from the \citet{Lamers05} model would then be one way to characterize the distributions. However, the variation of the disruption rate over space and time adds complexity to the shape of the age distribution that cannot be characterized by the \citet{Lamers05} model. Most importantly, the disruption time-scale systematically increases with age due to the mechanisms of cluster migration and natural selection (\citealt{Kruijssen11}, also see \citealt{Elmegreen10}). This has been named the `cruel cradle effect' \citep{Kruijssen12b} and is the immediate result of the fact that clusters are born in gas-rich and disruptive environments, which causes young clusters to be in more disruptive environments than older clusters. The impact of the cruel cradle effect is directly visible in Figure~\ref{fig:agedist} as a non-zero slope at young ages. This decline shows that (some) clusters get destroyed immediately at young ages, owing to the enhanced disruption rates and the variety of environments in which clusters reside.

Alternative ways to characterising age distributions exist. For instance, one could fit only the slope of the distribution at the old-age end, as was done in \citet{Kruijssen11}. However, this method is unreliable since the age distributions presented here often do not exhibit a single slope. It is therefore difficult to define over which range the slope should be measured. The third and final option is to take a statistical average such as the median or mean. Either of these quantities has the major advantage of being a non-parametric and model-independent diagnostic. Since the mean can be heavily biased by outliers, we decide to take the median. The median has the additional advantage that in the framework of the \cite{Lamers05} model, it can be easily related to the parameters $t_0$ and $\gamma$, enabling comparisons to previous work. We now turn to the quantification of this relation.

\subsection{Relating the Median Age to the Disruption Timescale}
\label{sec:derivation}
Here we present the justification for using the median age as a characterisation for the distribution as a whole. If we have a system that satisfies the assumptions of the \citet{Lamers05} model (i.e.~a disruption timescale that is constant throughout space and time), the median age can be simply expressed as combination of the variables $t_0$, $\gamma$, $\alpha$ and $M_{\rm min}$ (the minimum present-day cluster mass being considered). To demonstrate this relationship, we start with the joint age and mass distribution as derived by \cite{Lamers05}:
\begin{equation}
N(M,t) = \frac{S(t)}{\mu_{\rm ev}(t)^{1-\alpha}}\left(\frac{M}{M_{\odot}}\right)^{-\alpha}\left[1 + \frac{\gamma t}{t_0}\left(\frac{M}{M_{\odot}}\right)^{-\gamma}\right]^{(1-\alpha-\gamma)/\gamma} ,
\end{equation}
where $M$, $\alpha$ and $\gamma$ are as defined above (see Section \ref{sec:methods}), $S(t)$ is a constant proportional to the cluster formation rate, and $1 - \mu_{\rm ev}(t)$ is the fraction of cluster mass lost due to stellar evolution at age $t$.

Since we are only interested in the age distribution we integrate the joint distribution from the minimum present-day mass $M_{\rm min}$ to the maximum mass $M_{\rm max}$ to eliminate mass. For the purpose of simplifying the calculation, we take $M_{\rm max} = \infty$ and leave $M_{\rm min}$ as a variable. This assumption should have a neglible impact on the age distribution if the cluster mass function is sufficiently steep (e.g. $\alpha = 2$) since low mass clusters dominate the population in terms of number \citep{PortegiesZwart10}. After some rearranging of the integrand and the substitution $x = \left(M/\msun\right)^{\gamma}$, we obtain an integral that is simple to compute and then an expression for the age distribution:
\begin{equation}
N(t) = \frac{S(t)}{\alpha -1}\left[\frac{\gamma t}{t_0} +  \left(\frac{M_{\rm min}}{M_{\odot}}\right)^{\gamma}\right]^{(1-\alpha)/\gamma} ,
\end{equation}
which is valid for $1-\alpha/\gamma < 0$, because otherwise the integral does not converge. Assuming $\gamma > 0$, implying that more massive clusters have longer lifetimes as supported by theory and observations, we obtain a simpler restriction of $\alpha > 1$.  This too is consistent with the mass distributions of young stellar cluster populations \citep[e.g.][]{Larsen09,PortegiesZwart10,Kruijssen14}

Next, we normalize the distribution to create a probability distribution function. For the remainder of the calculation, we take the cluster formation rate $S(t)$ to be a constant, $S_0$. Integrating the distribution from $t = 0$ to $t = \infty$ and equating the result to unity, we obtain a normalization constant:
\begin{equation}
A = \frac{-t_0S_0}{\gamma(\alpha -1)}\left[\frac{\gamma t}{t_0} + \left(\frac{M_{\rm min}}{M_{\odot}}\right)^\gamma\right]^{(1 - \alpha + \gamma)/\gamma} ,
\end{equation}
with the restriction that $(1- \alpha - \gamma)/\gamma < 0$ or that $\alpha -1 > \gamma$ (again assuming that $\gamma > 0$). For $\alpha = 2$, we thus obtain $0<\gamma < 1$.

Next, we find the cumulative distribution function (CDF) by taking the integral of the normalized distribution function over the interval $t = [0, t']$, obtaining the following:
\begin{equation}
P(t<t') = 1 - \left[\frac{\gamma t'}{t_0}\left(\frac{M_{\rm min}}{M_{\odot}}\right)^{-\gamma} + 1 \right]^{(1 - \alpha + \gamma)/\gamma} .
\end{equation}
The final step is to set $P(t<t') = 1/2$ and solve for the age $t'$, thus giving the median age:
\begin{equation}
\label{eq:tmed}
t_{\rm med} = \frac{1}{\gamma}\left(2^{\gamma/(\alpha - \gamma-1)} - 1 \right)\left(\frac{M_{\rm min}}{M_{\odot}}\right)^{\gamma}t_0 ,
\end{equation}
which is valid for $\gamma > 0$ and $\alpha > 1$. If $\alpha =2$, as considered here, then the restriction becomes $0 < \gamma < 1$.  For a given $\gamma$ and $M_{\rm min}$, there exists a linear relationship between $t_{\rm med}$ and $t_0$. Hence, the median age of a group of clusters is an excellent probe for the disruption rate. Cluster populations with low median ages will have short disruption time-scales.

\begin{figure}
	\includegraphics[width=\hsize]{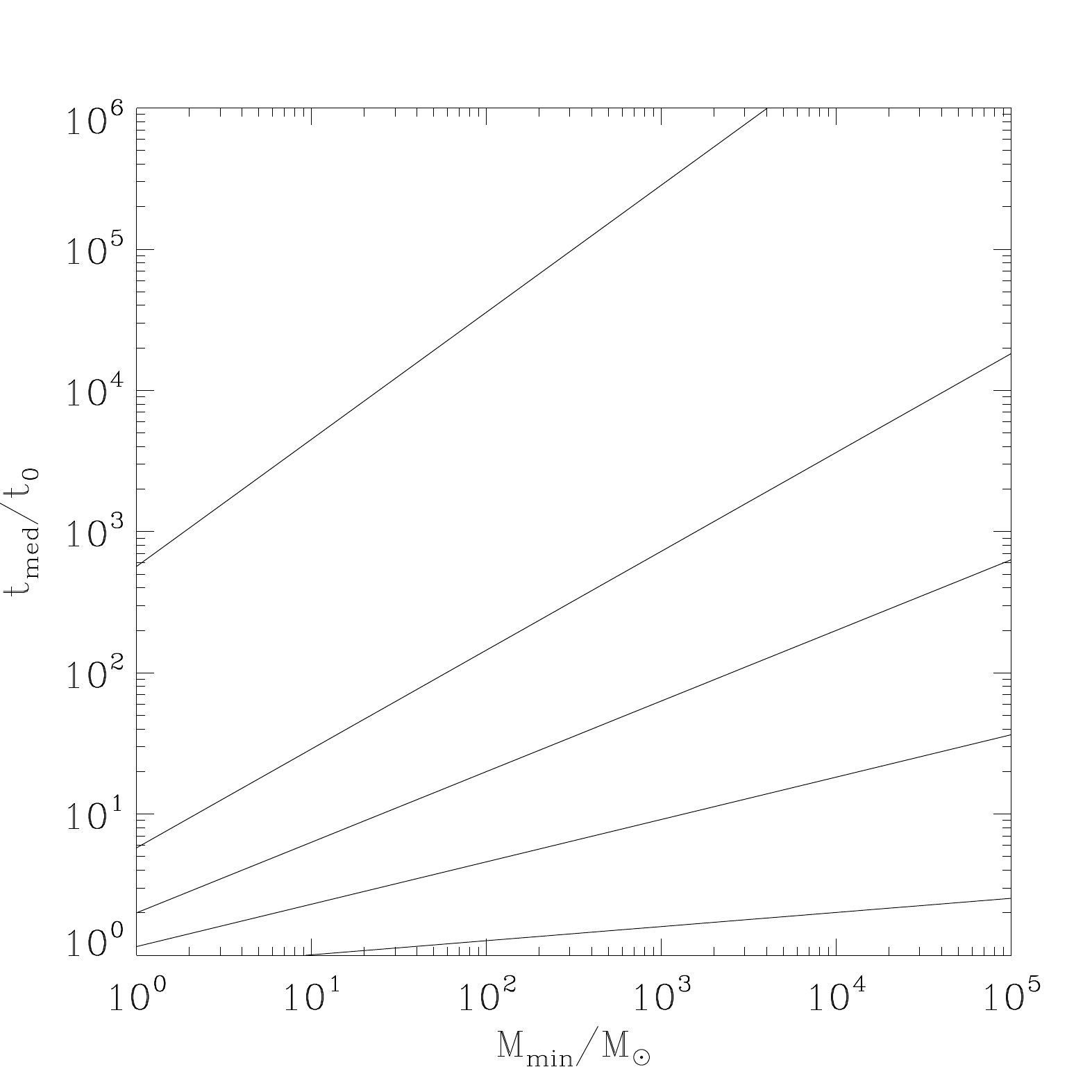}
    \caption{Ratio of the median cluster age and $t_0$ as a function of the minimum cluster mass, for different values of $\gamma$. From bottom to top we have $\gamma = [0.1, 0.3, 0.5, 0.7, 0.9]$. Equation~(\ref{eq:tmed}) shows that for fixed values of $M_{\rm min}$ and $\gamma$, the median cluster age is directly proportional to the disruption time-scale, implying that it can be used as a proxy for this disruption time-scale. The proportionality constant can be read off this figure.}
    \label{fig:tmedvst0}
\end{figure}
In Figure \ref{fig:tmedvst0}, we show the coefficient of the linear relation between the median age and the disruption time-scale parameter as a function of the minimum cluster mass, for different values of $\gamma$. For a given minimum mass, increasing gamma increases the ratio between $t_{\rm med}$ and $t_0$. The increase in $t_{\rm med}/t_0$, for a given $M_{\rm min}$ occurs because increasing $\gamma$ increases the ratio between the lifetimes of high mass clusters and low mass clusters and the presence of long lived massive clusters drives the median of the age distribution higher. Typical values of $\gamma$ are expected to lie in the range 0.6--0.8 when considering mass loss by two-body relaxation \citep{Lamers10}. However, the range of possible values increases when considering mass loss by tidal shocks, because the density-scaling of the corresponding disruption time-scale implies that the mass-radius relation plays a role in setting $\gamma$. Observations and theory suggest a weak mass-radius relation (resulting in $\gamma=0.6$--$0.7$, \citealt{Larsen04,Lamers05b,Gieles16}) and the value appropriate in the simulations is $\gamma\sim0.3$ for low-mass clusters and $\gamma\sim0.8$ for high-mass (weakly-disrupted) clusters in which the adiabatic correction is important (see Section~\ref{sec:methods}). These characteristic values hold for individual clusters, but the effective mass dependence of cluster disruption for the entire cluster population can be considerably lower if (some) tidal shocks are so strong that they can destroy clusters in a single encounter, independently of the cluster mass\footnote{For a given cluster radius, this mass-independence of cluster disruption applies up to some maximum mass, which can potentially exceed the mass of the most massive cluster. In that extreme case, disruption proceeds entirely independently of the mass.} \citep[also see Section~\ref{sec:median} below]{Kruijssen11}.

One important caveat to using $t_{\rm med}$ as a probe of $t_0$ in this work is that $\gamma$ is not necessarily the same at every point in the galaxy and therefore cannot be taken as a constant. For instance, we expect that different regions of the galaxy will have different velocity dispersions. The relative velocities between clusters and the gas will effect the duration of the encounters. Regions of the galaxy with low velocity dispersions will have longer encounters and hence the adiabatic correction (discussed in Section \ref{sec:methods}) will become more important. If the adiabatic correction is more significant, the mass dependence of the disruption time-scale goes up, i.e.~$\gamma$ is higher. Hence, regions of the galaxy with lower velocity dispersions have larger $\gamma$. However, the cluster mass functions in the different radial bins of Figure~\ref{fig:agedist} are almost identical, suggesting that $\gamma$ does not vary much as a function of radius (and environment) in the models considered. We therefore maintain that the median age is a suitable tracer for cluster disruption as long as the minimum present-day cluster mass is specified.

\begin{figure*}
	\includegraphics[width=0.49\hsize]{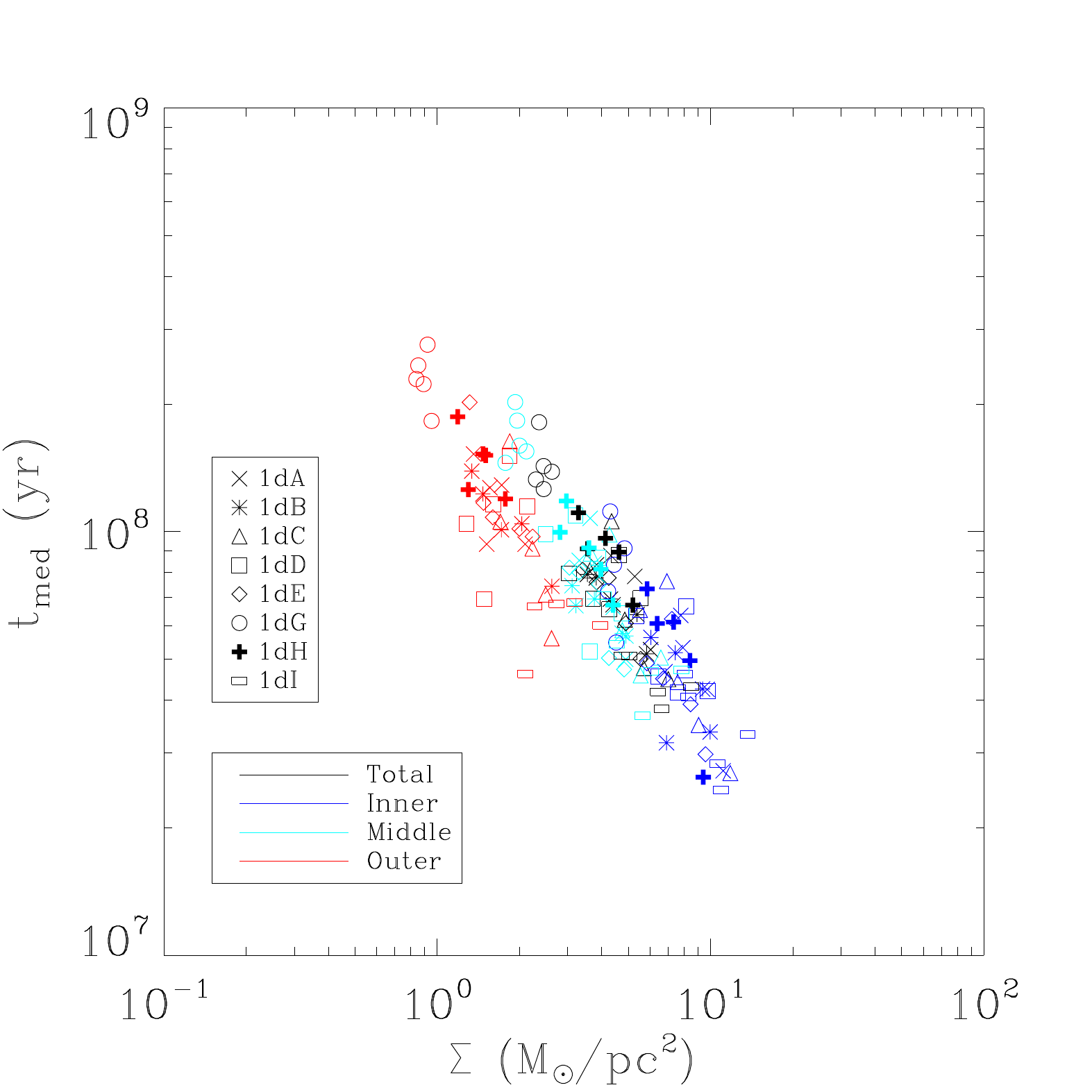}
\includegraphics[width=0.49\hsize]{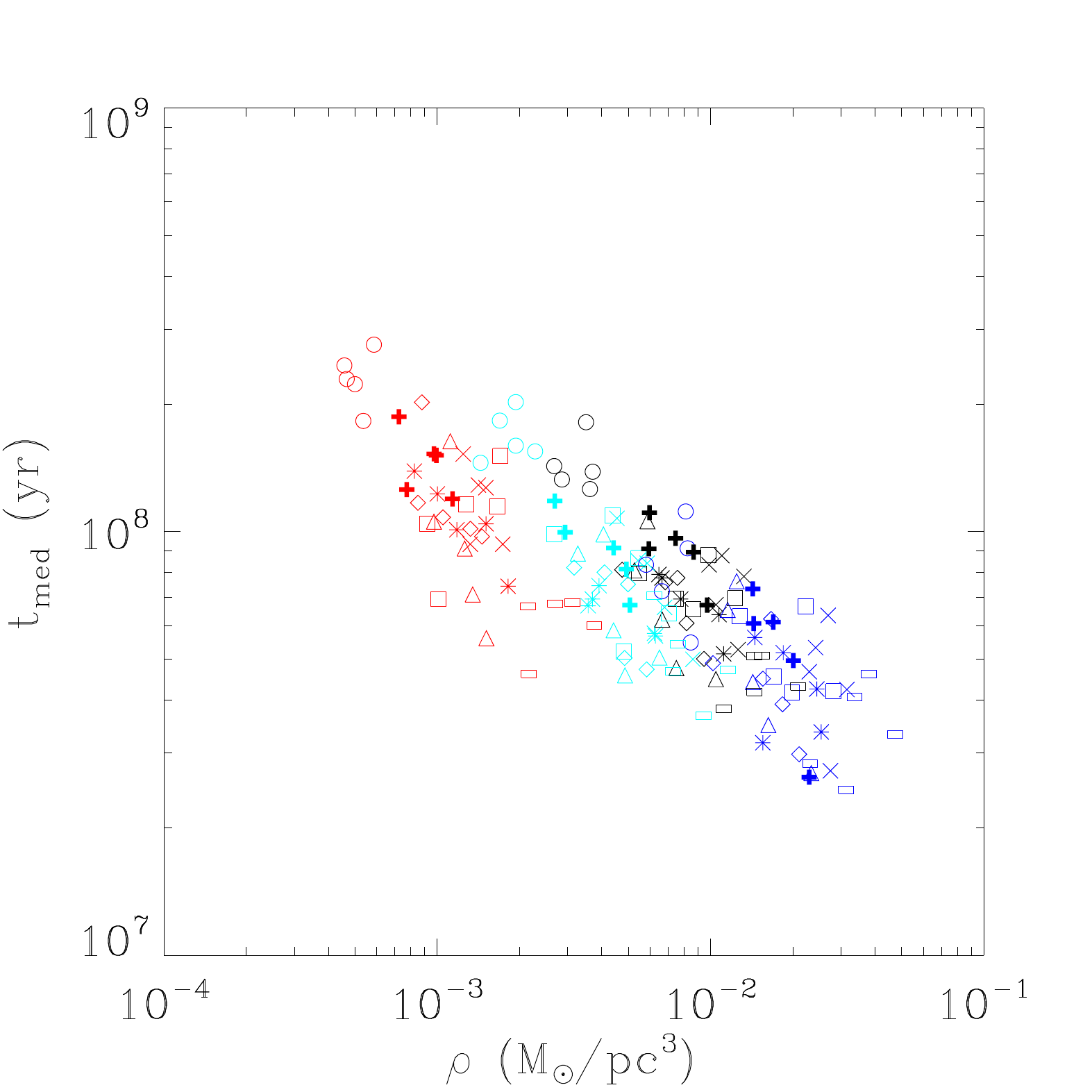}
\caption{Median cluster age of the entire cluster population ($M>10^2~\msun$) against gas surface density (left) and gas volume density (right). Each galaxy in the suite is represented by a different symbol while colours represent different radial bins, i.e.~inner (blue), middle (cyan), outer (red) and total (black).
\label{fig:medagesvsdens}}
\end{figure*}
Finally, the advantage of the linear relation between $t_{\rm med}$ and $t_0$ is that even in a composite cluster population with a range of disruption time-scales the median cluster age remains a good tracer of the population-averaged disruption time-scale. If we express the composite age distribution as a linear combination of different mono-disruption time-scale age distributions [$N(t,t_0)$], then the linearity of equation~(\ref{eq:tmed}) implies that the weighted average $t_0$ is related to the weighted average $t_{\rm med}$ in the same way as for a single disruption time-scale, provided that the populations associated to each individual mono-disruption time-scale age distribution have the same $\gamma$, $\alpha$, and $M_{\rm min}$.

\section{Correlation between the median age and the host galaxy properties} \label{sec:median}
We find that the median age correlates with many of the physical properties of the galaxy. In this section, we will describe some of these correlations and assess their physical significance. We consider eight isolated disc galaxy simulations from \citet{Kruijssen11,Kruijssen12} (see Table~\ref{tab:discs}),\footnote{We omit simulation 1dF because its total mass, star formation rate, and particle resolution differ by an order of magnitude from the other simulations. While \citet{Kruijssen11} carried out convergence tests and found only very weak systematic trends of the cluster destruction rate with resolution, simulation 1dF does exhibit differences relative to the other simulations that may be the result of resolution, galaxy mass, or otherwise. Because the present work focuses on a simple experiment in which we are probing the correlation between the cluster age distribution and the ISM properties, we choose to focus on those simulations where we can be certain that differences in galaxy mass, star formation rate, or resolution do not introduce any second-order trends. That way, we can isolate the dependence on the local ISM properties.} for each of which we consider five different snapshots at roughly 1~Gyr apart. This separation exceeds the median cluster age in all snapshots by a factor of 3--30, implying that these can be considered independent experiments. For each snapshot, we divide the galaxy disc into three separate radial bins that hold one third of all clusters in that snapshot, resulting in variable bin radii. Each variable of interest (i.e.~that is correlated with the median age) is calculated for the galaxy as a whole as well as in each radial bin. For most of the following analysis we use all of the clusters in each snapshot (i.e. no mass cut is made) unless otherwise stated. This implies a minimum cluster mass of $M_{\rm min} = 10^2~\msun$.

For each radial bin in each galaxy, in each of the 5 snapshots, we calculate the median cluster age as well as other properties that we might expect would influence the cluster disruption rate: the total gas surface density $\Sigma$, the mid-plane gas volume density $\rho$, the vertical velocity dispersion $\sigma_{\rm 1D}$, the median tidal heating parameter $I_{\rm tid}$, the angular frequency at that position in the galaxy $\Omega$, and the scale height $h$. This provides us with 120 independent experiments, as well as 40 linear combinations of these experiments that take the galactic average of each snapshot. The surface density is calculated by calculating the total gas mass in radial annuli in the face-on projection of each galaxy. The volume density is calculated by taking the surface density in that bin and dividing by twice the scale height. The scale height for each radial bin is calculated by fitting an exponential function to the vertical distribution of the gas in that bin. We calculate the vertical velocity dispersion around each star particle hosting clusters by locating all neighbouring gas particles within 2 kpc and calculating the dispersion of their velocities in the $z$ direction relative to that star particle. The average velocity dispersion for the population is then obtained by averaging over the star particles. This procedure mimics how velocity dispersions are measured observationally. The tidal heating parameter is saved for each cluster following Section~\ref{sec:methods} and the angular frequency is calculated from the rotation curve of each snapshot.

\begin{table*}
\centering
\begin{minipage}{12.6cm}
\centering
\begin{tabular}{|c|c|c|c|c|c|c|}

\hline
&&&&&&\\
& $\Sigma$ & $\rho$ & $\sigma_{\rm 1D}$ & $I_{\rm tid}$ & $\Omega$ & $h$ \\
&&&&&&\\
\hline

 &  \cellcolor{cyan!40}$\alpha_1 = -0.71$ & \cellcolor{green!70}$\alpha_1 = -1.28$ & \cellcolor{green!70}$\alpha_1 = -0.51$ & \cellcolor{cyan!40}$\alpha_1 = -0.61$ & \cellcolor{green!70}$\alpha_1 = -1.04$ & \cellcolor{green!70}$\alpha_1 = -0.94$ \\
$\Sigma$ & \cellcolor{cyan!40} & \cellcolor{green!70}$\alpha_2 = 0.34$ & \cellcolor{green!70}$\alpha_2 = -0.85$ & \cellcolor{cyan!40}$\alpha_2 = -0.10$ & \cellcolor{green!70}$\alpha_2 = 0.41$ & \cellcolor{green!70}$\alpha_2 = -0.33$ \\
  & \cellcolor{cyan!40} $\chi^2 = 627$ & \cellcolor{green!70}$\chi^2 = 557$ & \cellcolor{green!70}$\chi^2 = 561$ & \cellcolor{cyan!40}$\chi^2 = 619$ & \cellcolor{green!70}$\chi^2 = 573$ & \cellcolor{green!70}$\chi^2 = 557$ \\
\hline

 & &  \cellcolor{red!70}$\alpha_1 = -0.39$ & \cellcolor{cyan!40}$\alpha_1 = -0.21$ & \cellcolor{blue!50}$\alpha_1 = -0.19$ & \cellcolor{blue!50}$\alpha_1 = -0.86$ & \cellcolor{blue!50}$\alpha_1 = -0.94$ \\
$\rho$ &&\cellcolor{red!70} & \cellcolor{cyan!40}$\alpha_2 = -1.40$ & \cellcolor{blue!50}$\alpha_2 = -0.35$ & \cellcolor{blue!50}$\alpha_2 = 0.97$ & \cellcolor{blue!50}$\alpha_2 = -1.28$\\
  & & \cellcolor{red!70}$\chi^2 = 900$ & \cellcolor{cyan!40}$\chi^2 = 645$ & \cellcolor{blue!50}$\chi^2 = 825$ & \cellcolor{blue!50}$\chi^2 = 830$ & \cellcolor{blue!50}$\chi^2 = 830$ \\
 \hline

  &  &  &    \cellcolor{blue!50}$\alpha_1 = -2.49$ & \cellcolor{green!70}$\alpha_1 = -1.41$ & \cellcolor{cyan!40}$\alpha_1 = -1.59$ & \cellcolor{cyan!40}$\alpha_1 = -1.92$\\
$\sigma_{\rm 1D}$ &&& \cellcolor{blue!50}& \cellcolor{green!70}$\alpha_2 = -0.36$ & \cellcolor{cyan!40}$\alpha_2 = -0.37$ & \cellcolor{cyan!40}$\alpha_2 = 0.29$ \\
  & & &  \cellcolor{blue!50} $\chi^2 = 875$ & \cellcolor{green!70}$\chi^2 = 574$ & \cellcolor{cyan!40}$\chi^2 = 677$ & \cellcolor{cyan!40} $\chi^2 = 737$ \\
  \hline

  & & & &  \cellcolor{blue!50}$\alpha_1 = -0.64$ & \cellcolor{blue!50}$\alpha_1 = -0.53$ & \cellcolor{blue!50}$\alpha_1 = -0.79$ \\
$I_{\rm tid}$ &&&& \cellcolor{blue!50}& \cellcolor{blue!50}$\alpha_2 = -0.15$ & \cellcolor{blue!50}$\alpha_2 = -0.22$ \\
 & & & & \cellcolor{blue!50} $\chi^2 = 887$ & \cellcolor{blue!50} $\chi^2 = 878$ & \cellcolor{blue!50} $\chi^2 = 861$ \\
  \hline

   & &  & & & \cellcolor{red!70}$\alpha_1 = -0.79$ & \cellcolor{blue!50}$\alpha_1 = -1.88$ \\
$\Omega$ &&&&& \cellcolor{red!70}& \cellcolor{blue!50}$\alpha_2 = -1.23$\\
  & & & & & \cellcolor{red!70} $\chi^2 = 1064$ & \cellcolor{blue!50} $\chi^2 = 820$ \\
  \hline

   & & & & & & \cellcolor{red!70}$\alpha_1 = 0.77$ \\
$h$ &&&&&&\cellcolor{red!70}\\
  & & & & & & \cellcolor{red!70} $\chi^2 = 1503$ \\
 \hline

\end{tabular}
\label{tab:matrix}
\caption{Dependence of the median cluster age on seven different physical parameters, for single-parameter and two-parameter power-law fits. We list the best-fitting exponents and the goodness-of-fit statistic $\chi^2$ values, where $\alpha_1$ refers to the quantity listed in the left column and $\alpha_2$ refers to the quantity listed in the top row. The (off-diagonal) two-parameter fits follow the functional form of equation~(\ref{eq:fitform2}), whereas the (diagonal) single-parameter fits reflect equation~(\ref{eq:fitform1}). The cell colours reflect the goodness-of-fit, improving from red ($\chi^2\geq900$), blue ($750\leq\chi^2<900$), cyan ($600\leq\chi^2<750$), to green ($\chi^2<600$). The number of degrees of freedom is $N_{\rm deg}=118$ for the single-parameter fits, and $N_{\rm deg}=117$ for the double-parameter fits.}
\end{minipage}
\end{table*}

As shown in the previous section, we expect the median age to correlate with the various galaxy properties that determine $t_0$, the characteristic cluster disruption timescale. We first consider the correlation of cluster age with gas surface ($\Sigma$) and volume ($\rho$) density, in order to probe the interaction between clusters and the ISM. Figure \ref{fig:medagesvsdens} shows that the median age of the cluster populations in low density regions is higher then those in high density regions. This anti-correlation is no surprise, because tidal shocks by the gas are responsible for the bulk of the disruption in these simulations \citep[see Figure~8 of][]{Kruijssen11}. The density of the gas controls the frequency and strength of tidal shocks, leading to shorter cluster lifetimes and smaller median ages in higher-density environments.

Next to the clear first-order dependence on the gas density, each of the radial bins in Figure~\ref{fig:medagesvsdens} has its own correlation slightly shifted from the others (i.e.~each of the differently-coloured symbols has its own, offset trend). For a fixed gas density, clusters residing closer to the centre of the galaxy have higher median ages. This result suggests that there is a second important parameter that sets the amount of cluster disruption and determines the median age. We note that this shift from bin to bin cannot be accounted for by different bins having distinct $\gamma$. As discussed in Section \ref{sec:derivation}, $\gamma$ increases if the shock duration exceeds the dynamical time of a cluster. Given the lower angular frequencies and velocity dispersions at large radii, this implies that if $\gamma$ varies with the galactocentric radius, then the outer bins should have higher $\gamma$ and thus a larger ratio between $t_{\rm med}$ and $t_0$ rather than the other way around. Hence, if different $\gamma$ were contributing to the shift then we would expect the outer bins would be shifted up with respect to the inner bins, rather than down.

We also note that there are differences between the correlation with gas surface density and the correlation with gas volume density. To quantify these differences, we measure the slopes of these correlations and compute correlation coefficients. We find that both correlations are significant with correlation coefficients of -0.89 and -0.84 for surface density and volume density respectively. However, the slopes of the relations are different. We find that $t_{\rm med} \propto \Sigma^{-0.71}$ and $t_{\rm med} \propto \rho^{-0.39}$. However, given that these relations are weakened by the offsets between the different radial bins, we expect the fundamental dependence on $\Sigma$ and $\rho$ to be steeper.

The offset of each trend in each radial bin indicates that a second parameter is important. We also note that many of the galaxy properties are themselves correlated -- for example, both the gas density and the velocity dispersion decrease with galactocentric radius. Therefore, we perform a series of two-parameter fits of the form
\begin{equation}
\label{eq:fitform2}
t_{\rm med} = \alpha_0A^{\alpha_1} B^{\alpha_2} ,
\end{equation}
where $A$ and $B$ are two galaxy properties, and $\alpha_0$, $\alpha_1$, and $\alpha_2$ are free parameters. At the same time, we also look for the best fit for the single parameter fits 
\begin{equation}
\label{eq:fitform1}
t_{\rm med} = \alpha_0 A^{\alpha_1}
\end{equation}
for each galaxy property. To perform these fits, we use a weighted linear least squares algorithm, where the standard error in the median age for each data point is approximated as $\sigma_{t_{\rm med}}=1.253\sigma_t N_{\rm part}^{-1/2}$, with $\sigma_t$ the standard deviation of the cluster ages and $N_{\rm part}$ the number of star particles hosting clusters.\footnote{The dependence on the number of particles rather than the number of clusters is because all clusters in a single particle share the same age. The number of independent objects is therefore $N_{\rm part}$ and that number should be used to obtain the error on the mean or median.} This way, we obtain typical standard errors of $\sim5 \times 10^{-2}$ on $\log (t_{\rm med})$ for each data point. The resulting values of $\chi^2$ for each fit are shown in Table~\ref{tab:matrix}, where the boxes along the diagonal represent the one-parameter fits. We have three free parameters in our fit and 120 data points, so we expect a good fit to have $\chi^2$ values of a few hundred, as seen. 

\begin{figure}
\includegraphics[width=\hsize, trim = 0 20 0 0]{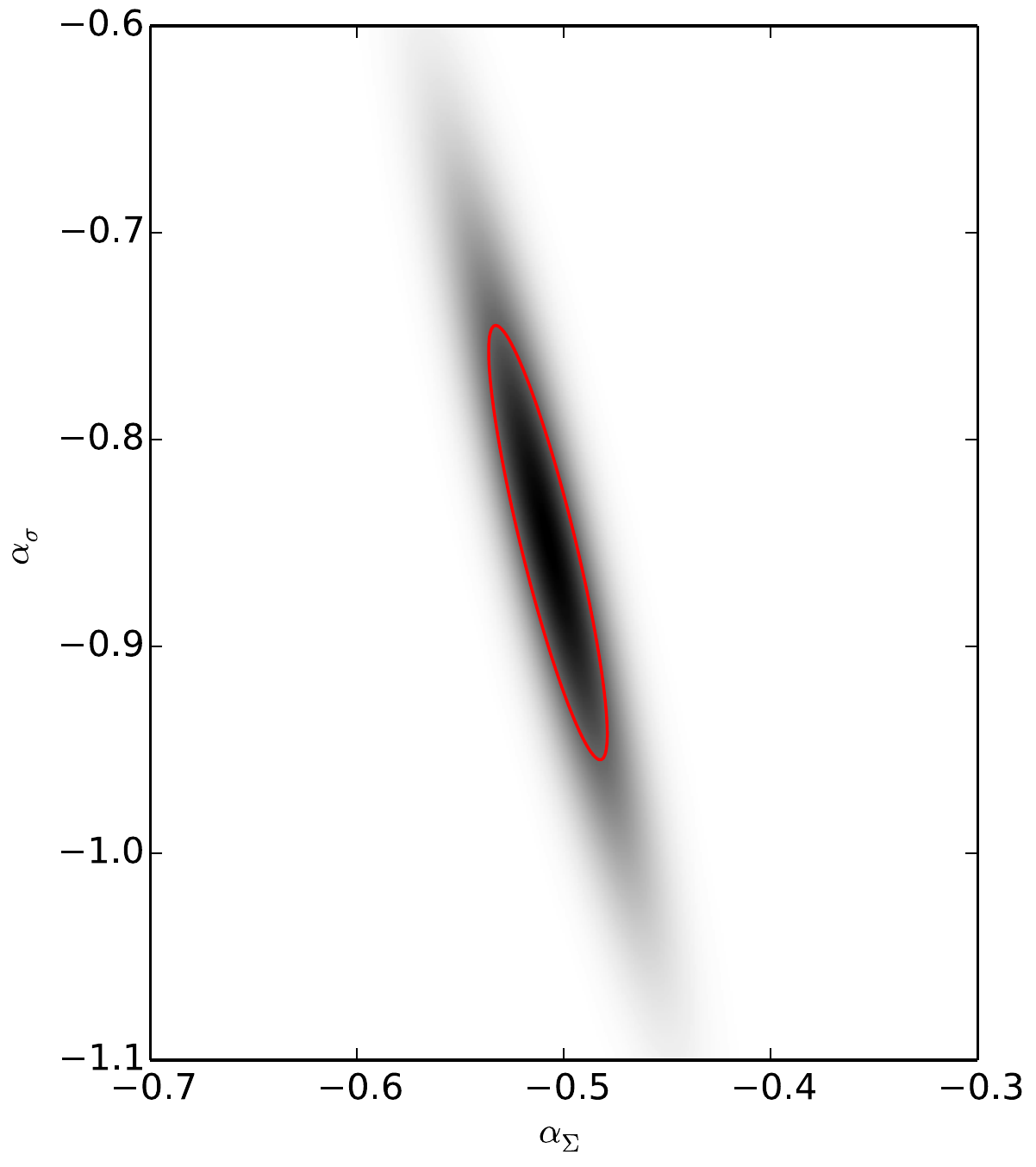}
\caption{Two-dimensional probability distribution function of the exponents of $\Sigma$ and $\sigma$ in equation~(\ref{eq:fit}). The $1\sigma$ uncertainty interval is shown as a red solid line.}
\label{fig:Landscape}
\end{figure}

The surface density $\Sigma$ is indeed the single parameter that predominantly controls the median cluster age. However, a better fit is found if we include both the surface density and another parameter, which can be either the volume density, the vertical velocity dispersion, or the scale height. The $\chi^2$ values for these three combinations are extremely similar and any of them would provide a good fit. Out of these options, we disregard the volume density because it is derived from the surface density and the scale height. Instead, we prefer the vertical velocity dispersion, because it is the most straightforward to obtain observationally \citep[e.g.][]{Leroy16} and it is the second-best of the single-parameter fits. The combination with the gas surface density gives a best fit of
\begin{equation}
\label{eq:fit}
t_{\rm med} =1.15~\gyr~\left(\frac{\Sigma}{\msun~\pc^{-2}}\right)^{-0.51 \pm 0.03} \left(\frac{\sigma_{\rm 1D}}{\kms}\right)^{-0.85 \pm 0.10} ,
\end{equation}
where we have included all clusters ($M>10^2~\msun$). The $1\sigma$ uncertainties on each of these exponents follow from the $\chi^2$ landscape around the best fit \citep{Press92}. We convert the $\chi^2$ landscape to a two-dimensional probability distribution function (PDF) by writing $P_{\rm 2D}\propto\exp{(-\chi^2/2)}$ and normalising the distribution to unity. The result is shown in Figure \ref{fig:Landscape}, highlighting that the exponent of $\sigma_{\rm 1D}$ is less well constrained than that of $\Sigma$, consistent with the fact that the single-parameter fits favour the gas surface density.

\begin{figure}
\includegraphics[width=1.04\hsize, trim = 0 20 0 0]{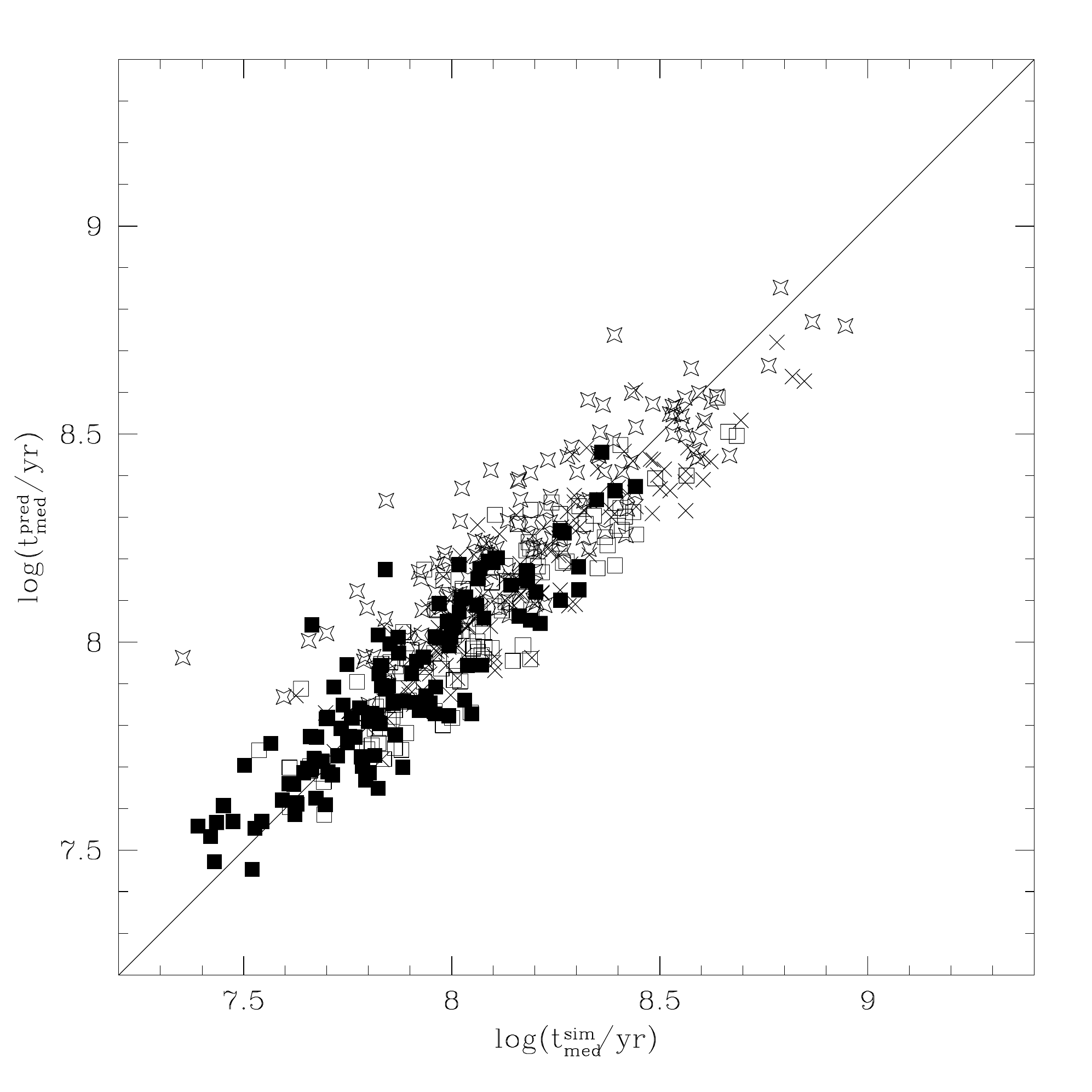}
\caption{Comparison of the proposed disruption law ($t_{\rm med}^{\rm pred}$, $y$-axis) with the median ages of the simulated cluster populations ($t_{\rm med}^{\rm sim}$, $x$-axis), for all considered snapshots. The different symbols indicate different minimum masses, with filled squares, open squares, crosses, and stars indicating $M>\{10^2, 10^3, 10^4, 10^5\}~\msun$, respectively.}
\label{fig:fitmass}
\end{figure}

Using the simulated cluster populations, we can independently fit for the dependence of the median age on the minimum present-day cluster mass, which according to equation~(\ref{eq:tmed}) provides the mass dependence of cluster disruption $\gamma$. We do this by measuring the logarithmic offset relative to equation~(\ref{eq:fit}) for minimum masses of $M_{\rm min}=\{10^2, 10^3, 10^4, 10^5\}~\msun$, which yields $\gamma\sim0.13$. This is considerably lower than the range of mass dependences $\gamma=0.3$--$0.8$ that is actually included in the simulations (see Section~\ref{sec:agedis}), which reflects the fact that clusters in the simulations are frequently destroyed by strong, single tidal shocks, independently of the cluster mass. Cluster populations in real galaxies may not show this same behaviour due to having a shallower mass-radius relation (see Section~\ref{sec:disc}). To account for the mass dependence of the median age, we introduce an extra term in equation~(\ref{eq:fit}) and scale it to typical surface densities and velocity dispersions, resulting in
\begin{equation}
\label{eq:fitmass}
\begin{aligned}
t_{\rm med} =&~92~\myr~\left(\frac{M_{\rm min}}{10^4~\msun}\right)^{0.13} \\
 & \times \left(\frac{\Sigma}{10~\msun~\pc^{-2}}\right)^{-0.51} \left(\frac{\sigma_{\rm 1D}}{10~\kms}\right)^{-0.85} .
\end{aligned}
\end{equation}
This fit is compared to the result of the simulation in Figure~\ref{fig:fitmass}, which demonstrates that the agreement is acceptable. The scatter around the 1:1 relation is roughly 0.1~dex.\footnote{In principle, a dependence of the median age on the minimum mass can also be accounted for by letting the slopes of the surface density ($\alpha_1$) and velocity dispersion ($\alpha_2$) vary as a function of mass, i.e.~$\alpha_1(M_{\rm min})$ and $\alpha_2(M_{\rm min})$, respectively. However, we stress that this (or any other) choice would not be as physically motivated as equation~(\ref{eq:fitmass}), which has the additional advantage of separating the dependence on environment ($\Sigma^{\alpha_1}$ and $\sigma^{\alpha_2}$) from the dependence on cluster properties ($M_{\rm min}^\gamma$). When performing a fit with minimum mass-dependent slopes, we find that the obtained median ages match the ones from equation~(\ref{eq:fitmass}) to well within 0.1 dex (i.e.~less than the scatter in Figure~\ref{fig:fitmass}). Therefore, the expression from equation~(\ref{eq:fitmass}) is strongly preferred.}

Any comparison of this relation to observations assumes that the mass dependence of cluster disruption is the same as in the simulations considered here. This is unlikely to be true, because the disruption time-scale due to tidal shocks depends on the cluster density and the adopted cluster mass-radius relation is steeper than observed. This leads to the excessive destruction of massive clusters and an underestimation of $\gamma$. Therefore, we generalise the relation below for the application to observed cluster populations.

\section{A semi-empirical cluster disruption law based on the host galaxy properties} \label{sec:disc}
In the previous sections, we have analysed self-consistent simulations of the formation and evolution of the cluster population in disc galaxies. We find that the median cluster age is strongly correlated with the local gas surface density and the local vertical velocity dispersion. Here we generalise the obtained relation to a semi-empirical disruption law for application to observations, carry out such a first comparison, and provide an initial interpretation and discussion of the caveats going into the disruption law.

\subsection{A generalised cluster disruption law} \label{sec:law}
The relation between the median cluster age and the gas surface density and velocity dispersion of equation~(\ref{eq:fit}) is derived directly from the simulations considered in this work. In the process, we have adopted certain physical descriptions for the cluster disruption process (Section~\ref{sec:methods}) and considered all clusters above $100~\msun$. Observations of cluster populations in real galaxies may deviate from these assumptions. The relation of equation~(\ref{eq:fit}) highlights the environmental dependence, but the dependence on cluster properties can be added through equation~(\ref{eq:tmed}). While $t_0$ is only set by the environment and therefore contains the scaling of equation~(\ref{eq:fit}), the dependence of the median cluster age on $\gamma$ and $M_{\rm min}$ should also be accounted for in equation~(\ref{eq:fit}). We define a parameter $\eta$ as
\begin{equation}
\label{eq:eta}
\eta = \frac{1}{\gamma}\left[2^{\gamma/(\alpha-\gamma-1)}-1\right] ,
\end{equation}
which ranges from $\eta=\ln{2}\approx0.69$ for $\lim_{\downarrow0}\gamma=0$ to $\eta=18.75$ for $\gamma=0.8$ (in both cases assuming that $\alpha=2$). We can combine this expression with equations~(\ref{eq:tmed}),~(\ref{eq:fit}) and~(\ref{eq:fitmass}) to obtain a generalised cluster disruption law for any value of $\gamma$. In order to do so, we have to choose at which minimum cluster mass to `pivot' the dependence on $\gamma$, i.e.~which mass to use as the fiducial `reference' value. The choice of `reference' mass comes down to the decision at which cluster mass the simulated disruption rate provides a good match to real-Universe disruption rates, irrespective of the value of $\gamma$. We adopt $M_{\rm min}=10^4~\msun$, because that is the mass-scale at which the mass-radius relation used in the simulations yields a radius that is very similar to the observed radii ($4.35~\pc$ versus the observed $3.75~\pc$, \citealt{Larsen04}), implying that the resulting disruption rate at that mass should be appropriate, irrespective of the value of $\gamma$. A simple substitution into equation~(\ref{eq:fitmass}) then yields\footnote{In obtaining the fit of equation~(\ref{eq:fitmass}), we have ignored the fact that the proportionality constant has a weak dependence on $\gamma$ too through $\eta$, as shown by equation~(\ref{eq:law}). However, this dependence only corresponds to a factor of 1.2 for the small values of $\gamma=0.1$--$0.2$ around the best-fitting value of $\gamma=0.13$ in Section~\ref{sec:median}, whereas the dependence on $M_{\rm min}^\gamma$ covers nearly a factor of 3 over the range of minimum masses considered.}
\begin{equation}
\label{eq:law}
\begin{aligned}
t_{\rm med} =&~255~\myr~\left[\frac{\eta(\alpha,\gamma)}{2.33}\right]\left(\frac{M_{\rm min}}{10^4~\msun}\right)^\gamma \\
 & \times \left(\frac{\Sigma}{10~\msun~\pc^{-2}}\right)^{-0.51} \left(\frac{\sigma_{\rm 1D}}{10~\kms}\right)^{-0.85} 
\end{aligned}
\end{equation}
where the normalisation of $\eta$ corresponds to the best match found in the comparison to observations in Section~\ref{sec:obs}, i.e.~$\gamma=0.54$ and $\eta=2.33$ (see below). However, any value of $\gamma$ can be substituted into this functional form. It is easy to verify that for $\gamma=0.13$, we retrieve equation~(\ref{eq:fitmass}).

We note that the quantitative predictions made with equation~(\ref{eq:law}) are quite sensitive to the choice of $\gamma$, because $t_{\rm med}/t_0=\eta(M_{\rm min}/\msun)^\gamma$. However, we cannot explicitly test this dependence using the simulations, because the value of $\gamma$ is an emergent property of the cluster disruption -- it is only slightly better constrained than in observations, because we know the adopted mass-radius relation. In principle, the mass-radius relation could be used to limit the range of possible values of $\gamma$ to $0.325\leq\gamma\leq0.8125$,\footnote{This is a range rather than a single value, because we do not know a priori if the tidal shocks are adiabatically dampened, which would change the mass dependence of cluster disruption.} but the possibility exists that some shocks are so extreme that clusters are destroyed independently of their masses, in which case the mass dependence can be as low as $0<\gamma<0.325$. As shown by the value $\gamma=0.13$ obtained for the simulations in Section~\ref{sec:median}, this indeed turns out to be the regime covered by a large number of models. When applying equation~(\ref{eq:law}) to observations, the correct value of $\gamma$ should be chosen by changing the lower cluster mass limit and considering the median ages of the resulting different sub-samples of clusters. As shown by equation~(\ref{eq:law}), a direct measurement of $\gamma$ appropriate for the cluster sample is obtained by fitting the power-law slope between the median age and the minimum present-day cluster mass.

\subsection{Comparison to observations} \label{sec:obs}
Across the nearby star-forming disc galaxy population, we have $\Sigma\sim10~\msun~\pc^{-2}$ \citep{Bigiel08} and $\sigma_{\rm 1D}\sim10~\kms$ \citep{Leroy16}, which is consistent with the default normalisation of equation~(\ref{eq:law}). For these characteristic numbers, we obtain a typical median age for observed cluster populations of several~$10^8~\yr$, which is in good qualitative agreement with ages reported in the literature \citep[e.g.][]{Hunter03,Bastian05,Bastian12,Fouesneau14,Adamo15b}.

\begin{table}
\centering
\begin{minipage}{\hsize}
\caption{Properties of observed cluster populations and their host galaxies}\label{tab:obs}
\begin{tabular}{c c c c c c}
\hline
Galaxy & $\Sigma$ & $\sigma_{\rm 1D}$ & $M_{\rm min}$ & $t_{\rm med}^{\rm obs}$ & $t_{\rm med}^{\rm pred}$ \\
 & $(\msun~\pc^{-2})$ & $(\kms)$ & $(\msun)$ & $(\yr)$ & $(\yr)$
\\\hline
M31 & $6$ & $7$ & $10^{3.0}$ & $10^{8.3}$ & $10^{8.1}$ \\
M51 & $80$ & $14$ & $10^{3.0}$ & $10^{7.3}$ & $10^{7.3}$ \\
M51 & $80$ & $14$ & $10^{4.0}$ & $10^{7.7}$ & $10^{7.8}$ \\
M51 & $80$ & $14$ & $10^{4.7}$ & $10^{8.2}$ & $10^{8.2}$ \\
M83i & $30$ & $15$ & $10^{3.7}$ & $10^{7.9}$ & $10^{7.9}$ \\
M83o & $7$ & $12$ & $10^{3.7}$ & $10^{8.3}$ & $10^{8.3}$ \\
\hline
\end{tabular}\\
Surface densities and velocity dispersions are taken from \citet[M31]{Braun09}, \citet{Leroy08} and \citet[M51]{Schuster07}, and \citet[M83]{Freeman17}. The cluster samples are taken from \citet[M31, with a median galactocentric radius of $R\sim12~\kpc$]{Fouesneau14}, \citet[M51, with a median galactocentric radius of $R\sim2.8~\kpc$]{Bastian05}, and \citet[M83]{Bastian12}. M83i refers to the inner field and M83o refers to the outer field observed by \citet{Bastian12}, which correspond to median galactocentric radii of $R=2.5~\kpc$ and $R=4.75~\kpc$, respectively. The uncertainties on $\Sigma$ and $\sigma$ are of the order 0.1~dex, resulting in uncertainties on $t_{\rm med}^{\rm pred}$ of $\sim0.15$~dex.
\end{minipage}
\end{table}
It is possible to make a closer comparison by considering the observed cluster (sub-)populations in individual galaxies or regions thereof. Table~\ref{tab:obs} summarises a small sample of comparisons for the well-studied cluster populations in the three galaxies M31, M51, and M83. The M51 sample is included with three different lower mass limits in order to test the dependence of equation~(\ref{eq:law}) on $M_{\rm min}$. M83 is divided into the inner and outer fields from \citet{Bastian12} to probe the environmental dependence in a homogeneous sample, although the inclusion of M31 already pushes the parameter space covered to extremely low gas surface densities.

Before deriving the predicted median ages, we require an estimate of $\gamma$. In Section~\ref{sec:law}, we proposed that the power-law slope between the median cluster age and the minimum present-day cluster mass provides a direct measurement of $\gamma$. We see that the median age in M51 increases by 0.9~dex over a 1.7~dex increase in $M_{\rm min}$ and carry out a power-law fit to the median age as a function of the minimum mass, resulting in a value of $\gamma\sim0.54$ for the cluster population in M51. For the shallow or absent mass-radius relations that are typically observed \citep{Larsen04,Scheepmaker07}, one would expect a value of $\gamma=0.7$--$1.0$ if disruption is driven by evaporation or impulsive tidal shocks.\footnote{For adiabatically-dampened shocks, this would increase even further up to $\gamma=1.75$--$2.5$.} The low best-fitting value of $\gamma$ indicates that the tidal shocks are largely impulsive and that a subset of them is so strong that they are capable of destroying clusters independently of their masses. We note that this value of $\gamma$ is consistent with the $\gamma=0.57\pm0.10$ obtained previously for M51 by assuming a constant disruption rate in time and space \citep{Lamers05b}. As discussed in this paper, our measurement of $\gamma=0.54$ is model-independent.

\begin{figure}
\includegraphics[width=\hsize, trim = 0 20 0 0]{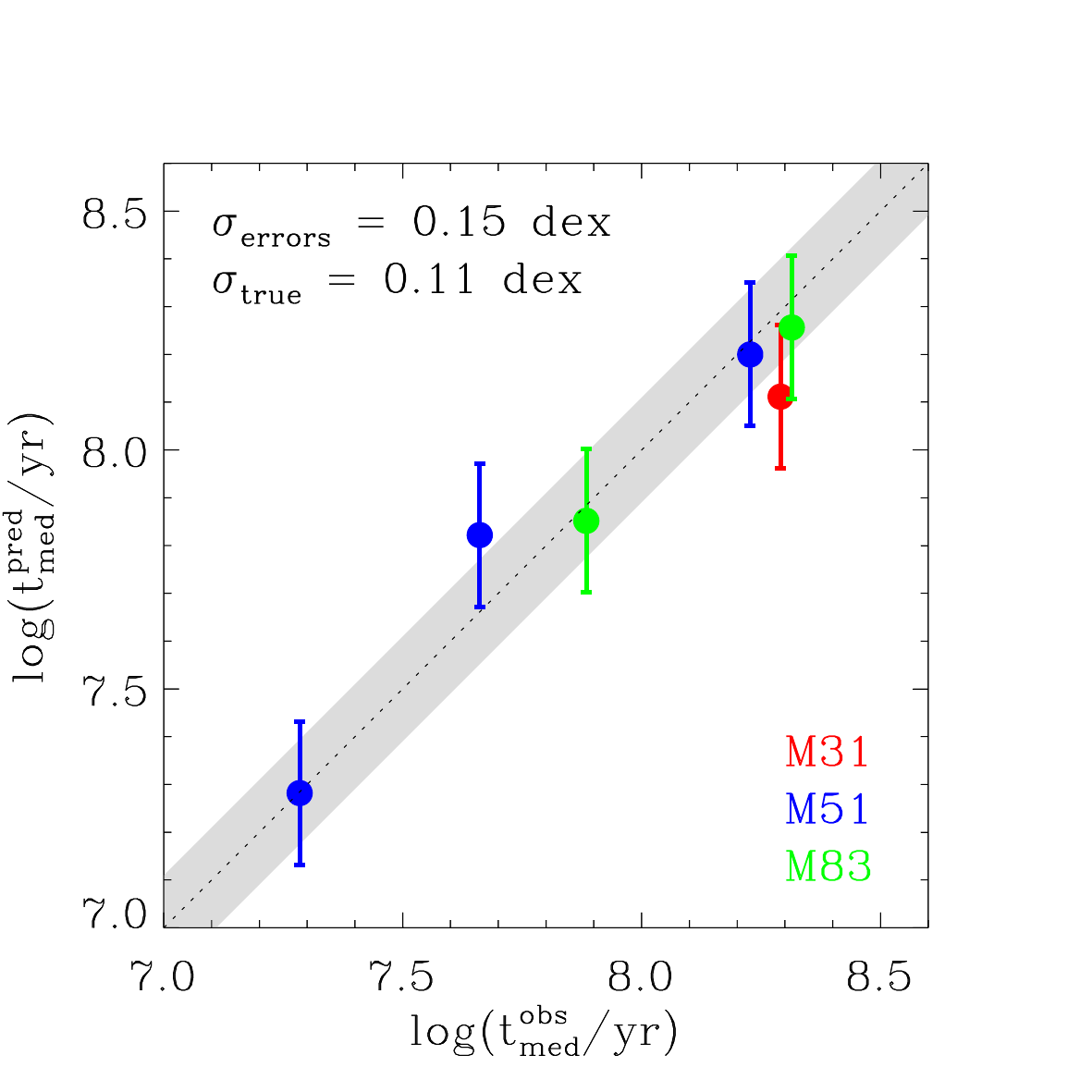}
\caption{Comparison between the observed median cluster ages from Table~\ref{tab:obs} and the median ages predicted by equation~(\ref{eq:law}), when adopting $\gamma=0.54$ as derived for M51. The colours refer to the different galaxies as indicated by the legend. The grey-shaded band indicates the standard deviation of the data points around the 1:1 relation. This scatter is 0.11~dex, similar to the approximate uncertainty on the predicted median ages of 0.15~dex.}
\label{fig:obs}
\end{figure}
We use the value of $\gamma=0.54$ derived for M51 to predict the median cluster ages in the final column of Table~\ref{tab:obs} for all six cluster samples. Figure~\ref{fig:obs} shows a comparison between the observed median cluster ages ($t_{\rm med}^{\rm obs}$) and the values predicted by the disruption law derived from our simulations ($t_{\rm med}^{\rm pred}$), exhibiting good agreement -- the numbers all differ by 0.2~dex or less and thus agree to within 60 per cent. This applies to median ages covering an order of magnitude, a similar range of gas surface densities, and gas velocity dispersions spanning a factor of 2. The scatter around the one-to-one agreement is 0.11~dex, quite similar to the approximate error bars on the predicted median ages (0.15~dex). This indicates that the proposed disruption law of equation~(\ref{eq:law}) provides a good description of the observed median ages.

Turning to M83, the magnitude of the environmental dependence is the same in the observations and in the theoretical relation, as the median cluster age drops by a factor of 2.5 when moving from the inner to the outer field. This is consistent with the analysis of \citet{Bastian12}, who find that for most mass-cuts, the outer field has a median age that is two to three times higher than the inner field. Finally, M31 is reproduced to within 60 per cent by equation~(\ref{eq:law}). We conclude that the proposed expression is accurate to well within a factor of 2 and thus provides a good first step in describing the complex interplay between cluster disruption and the ISM of the host galaxy.

\subsection{Interpretation, implications, and caveats}
We find that the median ages of stellar cluster populations most strongly correlate with the gas surface density and the vertical gas velocity dispersion, whereas they correlate less strongly with the gas volume density, the tidal heating parameter, and the angular frequency. Specifically, the median age $t_{\rm med}$ decreases with both $\Sigma$ and $\sigma$ and increase with the minimum cluster mass $M_{\rm min}$. Qualitatively, this is consistent with disruption by tidal interactions with the ISM in the regime where gravitational focussing is not important. Had gravitational focussing played an important role in driving cluster disruption and setting the median ages, then $t_{\rm med}$ should have increased with $\sigma$. A second reason why increasing the velocity dispersion could increase cluster lifetimes is that it puffs up the gas disc and decreases the mid-plane density. Again, no such effect is found. Instead, the dependence on the velocity dispersion (which admittedly is less strong that the correlation with the gas surface density) suggests that a high encounter rate caused by a high velocity dispersion is the main mechanism setting the median age at a given gas surface density. This is consistent with the fact that out of all dependences tested in Table~\ref{tab:matrix}, the tidal heating parameter provides some of the weaker correlations with the median age. The encounter rate thus outweighs the strength of each individual encounter. 

A key question to answer next is how the gas density dependence of equation~(\ref{eq:law}) obtained by fitting power laws to the output of the numerical simulations compares to that expected from analytic theoretical predictions for a single disruption rate ($t_{\rm med}\propto\rho^{-1.5}$). We cannot simply use the single-parameter fit to the gas volume density obtained above ($t_{\rm med}\propto\rho^{-0.39}$), because this relation may be affected by unidentified, second-order dependences. However, we can obtain the fundamental relation between the median cluster age and the gas volume density if we assume that the galaxy discs are in hydrostatic equilibrium, allowing us to write\footnote{We neglect the variation of the gas fraction $f_{\rm g}$, which formally should enter in a hydrostatic disc model. However, the effect is minor compared to the dynamic range of the variables under consideration (factor of $>10$, see Figure~\ref{fig:medagesvsdens}), because $\sigma\propto f_{\rm g}^{-1/2}$ \citep[e.g.][]{Krumholz05} and the gas fraction has a dynamic range of a factor of $<3$, implying a factor of $<1.7$ in $\sigma$.} $\rho\propto\Sigma^2/\sigma^2$ and $\sigma\propto (h\Sigma)^{1/2}$. At a fixed scale height, we have $\sigma\propto\Sigma^{1/2}$ and under this assumption equation~(\ref{eq:fit}) is equivalent to $t_{\rm med}\propto\rho^{-0.94}$. When considering the symbols of a single colour in the right-hand panel of Figure~\ref{fig:medagesvsdens}, which shows the relation between $t_{\rm med}$ and $\rho$, we see that a slope of $-0.94$ indeed describes the trend very well. The offset between the different colours is consistent with the radial increase of the scale height in the simulations. At small radii and large gas densities, the scale height is smaller than at large radii and low gas densities, causing the velocity dispersion to drop and the median age to increase. We thus conclude that the fundamental dependence of the median age on the gas density is of the order $t_{\rm med}\propto\rho^{-0.94}$ rather than the $t_{\rm med}\propto\rho^{-0.39}$ that is obtained with a single-parameter fit.

The fundamental scaling of the median age with the gas density ($t_{\rm med}\propto\rho^{-0.94}$) is still shallower than the analytical expectation of $t_{\rm med}\propto\rho^{-1.5}$, but this is to be expected. The analytical derivation holds for a cluster population that resides in single environment and does not undergo any systematic evolution with age. More realistic cluster populations such as the ones considered in the simulations reside in a range of environments that can be characterised by a distribution of disruption rates (and hence different intrinsic median cluster ages). As discussed in Section~\ref{sec:prelim}, such a cluster population is subject to the cruel cradle effect, in which the processes of cluster migration and natural selection lead to a decrease of the initial disruption rate with age. Towards older ages, this mechanism causes cluster samples in different environments to attain similar disruption time-scales, which weakens the dependence of the ensemble-averaged cluster disruption time-scale on the gas density in comparison to the density dependence of the instantaneous disruption time-scale of a single cluster. Therefore, older clusters no longer sample the mean ISM conditions in a galaxy, but are biased towards more benign conditions.

For the above reasons, the gas density dependence of the cluster lifetime for a range of disruption rates must be shallower than expected if the disruption rate were fixed in time and space. The amount by which the dependence is shallower is hard to predict analytically, because it depends on the complex interplay between the initial properties of the cluster population, the physics of cluster disruption, and the ISM structure. We therefore consider this to be an emergent effect of the numerical simulations and propose that a representative range of disruption rates leads to a flattening of the gas density dependence from $t_{\rm med}\propto\rho^{-1.5}$ to $t_{\rm med}\propto\rho^{-0.94}$. We stress that this does not mean that the fundamental physics of cluster disruption scale differently. When following an individual cluster in time, its instantaneous disruption time-scale may well scale as $\rho^{-1.5}$. It is only when considering the population statistics of the entire cluster sample that a systematic trend of disruption rate with age begins to emerge, leading to a weaker density scaling of the sample-averaged characteristic lifetime, in this case as $t_{\rm med}\propto\rho^{-0.94}$. A similar emergent effect of the cruel cradle effect on the slope of the age distribution was identified by \citet[Chapter 10.3]{Kruijssen11b}.

A final, important aspect of the analysis carried out in this paper is that we have not accounted for the environmental variation of the CFE. In the low gas surface density regime explored here, the CFE roughly scales as $\Gamma\propto\Sigma$ \citep{Kruijssen16,Johnson16}, meaning that one can expect variations of a factor several to an order of magnitude across the range of environments covered by the simulations. However, we do not expect that this affects any of our conclusions. As long as the CFE does not vary much in time (which should hold in isolated disc galaxies), any environmental dependence of the CFE only influences the normalisation of each individual age distribution and does not affect the shape of the distribution. In that case, the median age is therefore exclusively set by cluster disruption, we have shown in Section~\ref{sec:derivation}.

\section{Conclusions} \label{sec:concl}
In this paper, we have analysed self-consistent numerical simulations of the formation and evolution of the stellar cluster population in SPH simulations of isolated disc galaxies. The goal of this analysis has been to determine how the cluster age distribution is affected by cluster disruption, how the resulting age distribution depends on the galactic environment, and how we can use this dependence to probe the physics of cluster disruption. The conclusions of this work are as follows.
\begin{enumerate}
\item
The modelled age distributions slope downwards towards increasing ages even though the star formation rates in the simulations are roughly constant. This is indicative of cluster disruption, with 90 per cent of the clusters with initial masses $M>10^2~\msun$ being lost after $\sim2\times10^8~\yr$ and 99 per cent being lost after $\sim5\times10^8~\yr$ in our fiducial simulation. Cluster disruption is clearly environmentally dependent, as it proceeds more rapidly in the inner regions of the galaxy than in the outer regions.
\item
Contrary to previous work, these simulations include the variation of the disruption rate in time and space. As a result, no single functional form can fit all age distributions. We therefore propose to use a non-parametric and model-independent diagnostic of cluster disruption and suggest that the median age is the best quantity for this purpose. To facilitate comparisons to previous work, we derive the median age as a function of the disruption time-scale for mono-disruption time-scale cluster populations.
\item
We use a total of 40 independent simulation snapshots from our suite of isolated disc galaxy simulations and divide each of these into three radial bins with equal numbers of clusters, for a total of 120 cluster samples. We calculate the median cluster age and determine how it scales with a range of physical quantities by carrying out single- and two-parameter power law fits to the gas surface density, the mid-plane gas volume density, the one-dimensional velocity dispersion, the mean tidal heating parameter, the gas scale height, and the angular frequency.
\item
A two-parameter fit to the gas surface density and the one-dimensional velocity dispersion provides the best match with the least scatter. We find that the median age scales as $t_{\rm med}\propto\Sigma^{-0.51\pm0.03}\sigma_{\rm 1D}^{-0.85\pm0.10}$, which is qualitatively consistent with a disruption process that is dominated by tidal interactions with the ISM in the regime where disruption becomes more effective as the encounter rate increases with the velocity dispersion, i.e.~gravitational focussing is unimportant. We generalise this best-fitting relation to obtain a semi-empirical disruption law of that also includes the dependence of the median cluster age on the minimum cluster mass (equation~\ref{eq:law}).
\item
The scaling of the median cluster age enables the effective, population-averaged mass dependence of the cluster disruption time-scale to be measured as the power law slope between the median age and the minimum cluster mass of the sample. This is a model-independent way of measuring $\gamma$ from observed cluster samples, which is therefore preferred over previous approaches that assumed a constant disruption rate in time and space.
\item
We use the relation between the median cluster age and minimum present-day cluster mass in M51 to obtain a model-independent measurement of $\gamma=0.54$, which is broadly consistent with earlier measurements which assumed that the disruption rate is constant in time and space \citep[e.g.][]{Lamers05b}. We then compare the median cluster ages predicted by our disruption law to those of observed cluster populations in M31, M51 (for three different minimum cluster masses), and M83 (at two different galactocentric radii). The predictions and observations agree to within 60 per cent or less in all six samples considered.
\item
We show that the proposed disruption law implies a cluster sample-averaged scaling of the median cluster age with the mid-plane gas volume density of $t_{\rm med}\propto\rho^{-0.94}$, which is shallower than the prediction from analytic theory of $t_{\rm med}\propto\rho^{-1.5}$. This difference arises due to the cruel cradle effect, in which the disruption time-scale systematically increases with cluster age due to cluster migration (by galactic dynamics) and natural selection (i.e.~the survival of clusters in benign environments). Towards older ages, this mechanism causes cluster samples in different environments to attain similar disruption time-scales, which weakens the dependence of the ensemble-averaged cluster disruption time-scale on the gas density in comparison to the density dependence of the instantaneous disruption time-scale of a single cluster.
\end{enumerate}
These results show that a systematic analysis of the emergent demographics of cluster populations in disc galaxy simulations reveals useful trends that probe the variation of the cluster disruption rate in time and space. Our approach of quantifying the environmental variation of cluster age distributions provides a step forward relative to previous work, in which the disruption rate of a single cluster population was assumed to be constant, because our use of the median cluster age is appropriate even if the cluster population has a spread of disruption rates. It is encouraging that the semi-empirical disruption law derived here provides a good match to the small set of observed median cluster ages considered. We encourage observational studies to carry out similar comparisons, with a focus on varying the minimum present-day cluster mass to obtain the mass dependence of cluster disruption $\gamma$ and on using ALMA to probe a wide range of gas surface densities and velocity dispersions.

This paper builds on previous work to provide a natural step towards understanding the complex interplay between cluster formation, cluster disruption, and the ISM. A theoretical understanding of this interplay is necessary to realise the potential of stellar clusters as tracers of galaxy formation and evolution. Future work is needed to improve on the numerical methods used and the idealised nature of the simulated galaxies. Modern simulations with cosmologically-motivated boundary conditions will set the next step in probing the co-evolution of cluster populations and galaxies.

\section*{Acknowledgements}
We thank the anonymous referee for a helpful report. The work presented here was funded by the Michael Smith Foreign Study Supplement (MSFSS) administered by the National Sciences and Engineering Research Council of Canada (NSERC). MM and AS acknowledge additional funding from NSERC. JMDK gratefully acknowledges funding from the German Research Foundation (DFG) in the form of an Emmy Noether Research Group (grant number KR4801/1-1, PI Kruijssen), and from the European Research Council (ERC) under the European Union's Horizon 2020 research and innovation programme via the ERC Starting Grant MUSTANG (grant agreement number 714907, PI Kruijssen). MM extends many thanks to JMDK and the rest of the MUSTANG group at the Astronomisches Rechen-Institut (ARI) at Heidelberg University, where the majority of this work was carried out, for providing a pleasant and productive working environment.

%%%%%%%%%%%%%%%%%%%% REFERENCES %%%%%%%%%%%%%%%%%%

\bibliographystyle{mnras}
\bibliography{mybib}

\bsp
\label{lastpage}
\end{document}